\documentclass[lettersize,journal]{IEEEtran}
\usepackage{amsmath,amsfonts}
\usepackage{algorithmic}
\usepackage{algorithm}
\usepackage{array}
\usepackage[caption=false,font=normalsize,labelfont=sf,textfont=sf]{subfig}
\usepackage{textcomp}
\usepackage{stfloats}
\usepackage{url}
\usepackage{verbatim}
\usepackage{graphicx}
\usepackage{cite}
\usepackage{bm}
\usepackage{amssymb}
\usepackage{booktabs}
\usepackage{overpic}
\usepackage{multirow}
\usepackage{makecell}
\usepackage{hyperref}
\usepackage{xcolor}
\begin{document}

\title{Single Camera based Simulation and Try-on Ready Avatar Modeling}

\author{Honghu~Chen\textsuperscript{*},
        Bo~Peng\textsuperscript{*},
        Jun~Xiang,
        Jingtao~Zhou,
        Xujing~Chen,
        and~Juyong~Zhang$^\dagger$,~\IEEEmembership{Member,~IEEE,}% <-this % stops a space

\IEEEcompsocitemizethanks{\IEEEcompsocthanksitem  H. Chen and X. Chen are with the School
        of Artificial Intelligence and Data Science, University of Science and Technology of China.}
\IEEEcompsocitemizethanks{\IEEEcompsocthanksitem  B. Peng, J. Xiang, J. Zhou and J. Zhang are with the School
of Mathematical Science, University of Science and Technology of China.
}
\IEEEcompsocitemizethanks{\IEEEcompsocthanksitem  {*}
These two authors contributed equally to this work.
}
\IEEEcompsocitemizethanks{\IEEEcompsocthanksitem  $^\dagger$Corresponding author. Email: \texttt{juyong@ustc.edu.cn}.

© 2026 IEEE. Personal use of this material is permitted. Permission from IEEE must be obtained for all other uses, in any current or future media, including reprinting/republishing this material for advertising or promotional purposes, creating new collective works, for resale or redistribution to servers or lists, or reuse of any copyrighted component of this work in other works.

}

}

        % <-this % stops a space

\markboth{ IEEE TRANSACTIONS ON VISUALIZATION AND COMPUTER GRAPHICS}%
{Shell \MakeLowercase{\textit{et al.}}: A Sample Article Using IEEEtran.cls for IEEE Journals}


\maketitle

\begin{abstract}
Reconstructing and rendering high-fidelity, animatable, and try-on ready virtual avatars using only a single camera is crucial for many applications.
However, existing try-on-ready human avatar reconstruction methods struggle to achieve this from monocular videos or single images,
primarily due to two issues:
(1) Existing approaches typically use a single layer to represent both garments and the underlying body simultaneously, ignoring their distinct motion characteristics; 
(2) Input captured from a single camera often fails to fully cover both garments and the underlying body.
To address these challenges, we propose a physically integrated double-layer GS-Mesh 3D clothed human representation, reconstructing decoupled clothed avatars from monocular videos or even single images by incorporating 2D generated data. 
While 2D generation methods can produce video sequences from given pose sequences, they struggle to maintain inter-frame consistency. 
We further decompose the inter-frame inconsistencies in generated data into geometric and appearance inconsistencies. 
Based on the 3D representation, we ensure the final reconstruction and rendering results are both 3D-consistent and temporally coherent.
Compared to existing 3D garment transfer methods relying on multi-view reconstruction, or most 2D approaches that use diffusion models for image-based garment editing without novel view/pose synthesis, our method only needs monocular video or even a single image as input while achieving \textcolor{black}{visually plausible cloth dynamics} effects during virtual try-on and novel pose animation. 
This represents a significant advancement in achieving \textcolor{black}{virtual try-on with visually plausible cloth dynamics} with just a single camera.
\end{abstract}

\begin{IEEEkeywords}
physical simulation, reconstruction, virtual try-on.
\end{IEEEkeywords}

\begin{figure*}[ht]
  \centering
  \includegraphics[width=\textwidth]{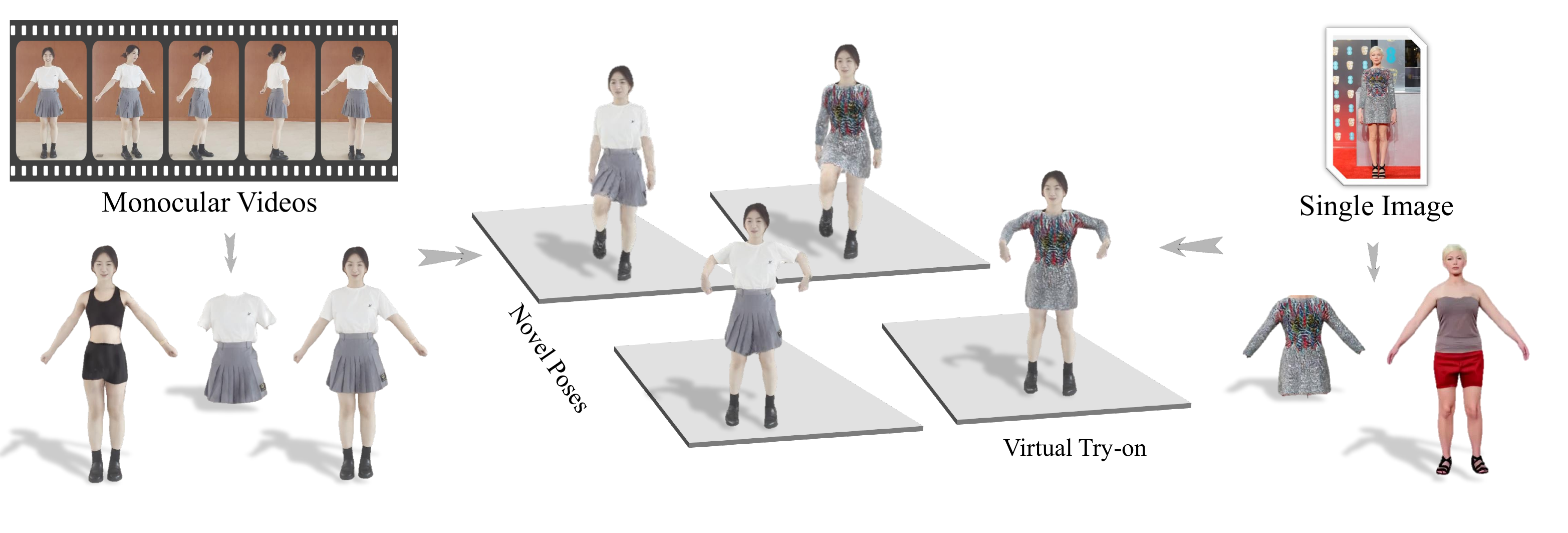}
  \vspace{-13mm}
  \caption{Given a monocular input (a monocular video or a single image), our method reconstructs a disentangled clothed human avatar. The reconstructed avatar supports \textcolor{black}{virtual try-on through physical simulation and produces visually plausible cloth dynamics}.}
  % \caption{In the pose space, we use three latent spaces to encode query points in canonical pose. First, First, calculate the pose dependent offset for all query points. Then transfer them to the posed space by neural LBS.}
  % \label{arch}
  \end{figure*}

% \vspace{-3mm}
\section{Introduction}
\IEEEPARstart{H}{igh-fidelity} 
digital humans are widely applied in fields such as VR/AR and the metaverse.
Compared to cinematic-quality digital human reconstructions that require a large number of cameras, high-fidelity digital humans that can be reconstructed at low cost are more suitable for widespread promotion and mass applications.
Existing monocular reconstruction methods mainly focus on the appearance of digital humans and represent the clothed human as a single layer. 
This results in significant challenges for the human avatar when changing clothing or encountering unseen poses.
% This makes it difficult for clothed human avatars to be directly applied to applications such as virtual try-on.
% Virtual Try-On uses computer vision, 3D modeling, and augmented reality (AR) technologies to allow users to try on clothing in a virtual environment without physically wearing them. It is widely used in e-commerce to help users make better choices while shopping online.
% An ideal virtual try-on should rely solely on easily accessible single-view data, accurately presenting the clothing's physical state based on the underlying body shape and pose, while delivering realistic 3D effects.

We aim to develop a method to reconstruct decoupled human avatars from monocular videos or single images, and \textcolor{black}{enable virtual try-on through physical simulation with visually plausible cloth dynamics}.
However, this is a highly challenging task due to the following reasons: 1) Monocular inputs only provide single-view 2D image information, with unobservable regions existing making it difficult to reconstruct the geometry and appearance of decoupled clothing and body; 2) When utilizing 2D generated data to supplement missing information in monocular inputs, temporal inconsistency issues arise; 3) In virtual try-on or animation production, ensuring the reliability of rendering results in terms of 3D consistency, appearance, and \textcolor{black}{visually plausible cloth dynamics is crucial}.

To address these challenges, we propose a novel method for reconstructing a double-layered clothed human avatar from monocular inputs, which integrates a physical simulator to \textcolor{black}{generate novel-pose animations with visually plausible cloth dynamics}. 
It also combines generative priors to ensure high rendering quality when presenting novel clothing and novel poses.
When generalizing to challenging poses or clothing, unseen regions 
% (such as unseen clothing areas or body parts hidden by clothing in the input images) 
may produce artifacts. 
This problem is overlooked by most existing monocular human avatar reconstruction methods~\cite{peng2023intrinsicngp, peng2021animatable, Lei_2024_CVPR, wen2024gomavatar, pramishp2024iHuman}.
To solve this problem, we propose a strategy that combines generation and reconstruction, using the generated image information to guide the reconstruction of unseen regions in the input images.
However, generated image sequences often suffer from inconsistency with real data as well as inter-frame inconsistency within the generated data.
We address this issue by modeling inconsistencies as geometric and appearance inconsistencies.
Geometric inconsistencies include inconsistencies in human pose and geometric shape. We design isolated deformation representations for real and \textcolor{black}{generated} data and perform pose re-tracking on them. Appearance inconsistencies include lighting and subtle style inconsistencies, which we align using SH coefficients and perceptual loss.
Using this strategy, we can significantly reduce the impact of inconsistency on the quality of generated sequence reconstructions.

With the monocular inputs and generated image sequences, our reconstructed clothed human avatar demonstrates strong generalization capabilities for novel poses and novel clothing.
Compared to existing GS-based monocular reconstruction methods with single-layer representation, our method shows strong generalization ability to novel poses. 
Without manual intervention, we directly apply it to virtual try-on and demonstrate its effectiveness.
Unlike existing 2D try-on methods that typically fail to generalize to novel poses or views, and 3D try-on methods that require multi-view inputs, our method can reconstruct human avatars wearing different clothing from only monocular inputs and \textcolor{black}{visually plausible cloth dynamics for novel poses and garments through physical simulation}.
In application experiments, we demonstrate \textcolor{black}{virtual try-on animations driven by physical simulation with visually plausible cloth dynamics}.
% Existing try-on methods can be categorized into 2D-based and 3D-based approaches.
% 2D-based try-on~\cite{han2017viton, wang2018toward, ge2021parser, xie2023gpvton, xu2024ootdiffusion, kim2024stableviton, morelli2023ladi, Zhu_2023_CVPR_tryondiffusion} primarily leverages generative models, which can produce high-quality images but fail to ensure 3D consistency and physical realism.
% 3D-based try-on methods mainly include two types: those~\cite{cao2024gsvton, chen2024gaussianvton, DreamVTON} using 2D priors to guide 3D optimization, ensuring limited 3D consistency for novel views and motions; and those~\cite{lin2024layga, PhysAavatar24, peng2024pica} reconstructing layered clothing and body representations, enabling better results through clothing transfer but often requiring multi-view inputs.
% 3D-based try-on methods~\cite{lin2024layga, PhysAavatar24, peng2024pica} reconstruct layered clothing and body representations, achieve better results through clothing transfer, but often require multi-view inputs.
In summary, the contributions of this paper include the following aspects:
\begin{itemize}
\item We propose a method to reconstruct a disentangled clothed human, \textcolor{black}{including the garment and an underlying-body representation whose unobserved regions are plausibly completed}, from monocular inputs.
\item To overcome the inter-frame inconsistency in generated data, we separately address its geometric inconsistency and appearance inconsistency to mitigate their impact on reconstruction quality.
\item Combining our carefully designed double-layer Mesh-GS and physical simulations, the reconstructed clothed human avatars can \textcolor{black}{produce visually plausible cloth dynamics and high-fidelity rendering results for virtual try-on}.
\end{itemize}
\section{Related Work}

% Head 2

\subsection{Reconstruction from Monocular Input.}

\textbf{Reconstruction from videos.}
Recently, several methods based on mesh representations, implicit representations, and Gaussian splatting have focused on reconstructing the geometry and appearance of clothed human bodies from monocular videos.

Mesh-based methods use parametric human~\cite{loper2015smpl, Anguelov05scape:shape, jiang2020bcnet, SMPLX_2019} and clothing models to obtain reconstruction results through regression~\cite{Kanazawa_2018_CVPR, pavlakos2018learning, omran2018neural} or optimization~\cite{bogo2016keep}, but these methods generally lack details. Other methods~\cite{alldieck2018video, grassal2021neural, li2023tcsvt} rely on parametric models or specially designed templates and optimize vertex offsets to achieve detailed reconstruction, but the templates are often not directly applicable to diverse categories of clothing.
Implicit function-based reconstruction methods use neural networks to model occupancy fields~\cite{saito2019pifu, saito2020pifuhd}, signed distance fields~\cite{Chen2024NeuralABC, jiang2022selfrecon}, and density fields~\cite{peng2021animatable, peng2023intrinsicngp} to represent 3D human or render fine details. 
These methods can represent various types of clothing without topological constraints.
However, these methods typically have long rendering times and limited editability.
3D Gaussian Splatting (3DGS)-based reconstruction methods~\cite{kerbl3Dgaussians} greatly improve rendering speed and editability compared to NeRF-based approaches~\cite{mildenhall2020nerf}. Some methods extend 3DGS to dynamic human bodies~\cite{Lei_2024_CVPR, wen2024gomavatar, pramishp2024iHuman} using SMPL-guided deformations, enabling faster generation of editable animated human models.

\textbf{Reconstruction from few-shot images.}
Reconstructing avatars from few-shot images remains a highly challenging task due to the sparsity of available input. PIFu~\cite{saito2019pifu} and following works~\cite{saito2020pifuhd,xiu2022icon,xiu2023econ,corona2023structured,zheng2021pamir} introduce pixel-aligned features to facilitate novel view synthesis. SHERF~\cite{hu2023sherf} learns a generalizable human NeRF but suffers from slow rendering speeds, while GAS~\cite{lu2025gas} further incorporates a video diffusion model to enhance output sharpness.
A prominent direction in recent work is to utilize diffusion priors to compensate for missing information. This includes generating novel views to enrich the input~\cite{li2024pshuman, liu2024human, zhang2024sifu, qiu2024AniGS}, and applying Score Distillation Sampling (SDS)~\cite{poole2022dreamfusion} to generate 3D avatars from 2D priors~\cite{huang2024tech, xiu2024puzzleavatar, yang2024have, zhang2024humanref, wang2025wonderhuman}.
\textcolor{black}{Although AniGS~\cite{qiu2024AniGS} also leverages generated observations for single-image avatar reconstruction, it formulates cross-view inconsistencies as temporal variations and handles them using 4DGS. In contrast, our method targets a disentangled, try-on-ready body-garment representation. We separately generate key-pose clothed-human observations and underlying-body observations, and explicitly decompose generation errors into geometric and appearance inconsistencies. In addition, our representation supports garment transfer and subsequent cloth simulation rather than skinning-based avatar animation alone.}
IDOL~\cite{zhuang2024idolinstant}, Vid2Avatar-Pro~\cite{guo2025vid2avatarpro} and LHM~\cite{qiu2025LHM} perform single-image human reconstruction efficiently in a feed-forward manner, but often produce coarse results when handling custom inputs.
However, most of these methods treat the clothed avatar as a whole.
In contrast, our method employs a double-layer representation and leverages 2D diffusion priors for both the clothed avatar and the underlying body, which is necessary for clothing transfer.

\subsection{Virtual Try-on}

\textbf{2D Virtual Try-on.}
Image-based virtual try-on aims to generate composite images of a person wearing a specified garment. Deformation-based methods like VITON~\cite{han2017viton} and CP-VTON~\cite{wang2018toward} use Thin Plate Spline (TPS)~\cite{bookstein1989principal} transformation and a staged approach, while PF-AFN~\cite{ge2021parser} combines knowledge distillation, and GP-VTON~\cite{xie2023gpvton} improves detail consistency through local flow, but alignment issues remain. 
Diffusion-based methods overcome these limitations by leveraging pre-trained models. WarpDiffusion~\cite{zhang2023warpdiffusionefficientdiffusionmodel} bridges deformation and diffusion with local feature attention, TryOnDiffusion ~\cite{Zhu_2023_CVPR_tryondiffusion} uses UNet for feature extraction, LaDIVTON ~\cite{morelli2023ladi} maps features to CLIP~\cite{radford2021learning}, StableVITON~\cite{kim2024stableviton} introduces a zero-cross attention block, and OOTDiffusion~\cite{xu2024ootdiffusion} fine-tunes the garment UNet, further improving performance.

Image-based methods can produce high-quality images, but they struggle to achieve good results across diverse poses and to generate continuous and consistent results for sequences.

\textbf{3D Virtual Try-on.} The research on 3D virtual try-on mainly focuses on modeling layered clothing and bodies, studying how clothing deforms under the influence of the underlying body's shape and motion, or how it behaves according to physical laws.
These methods~\cite{li2023isp, santesteban2022snug, bertiche2020pbns} can model the dynamic deformation of clothing in great detail, but typically at the geometric level, they are unable to achieve high-fidelity rendering quality.
Diffusion-based methods~\cite{cao2024gsvton, chen2024gaussianvton, DreamVTON} use differentiable 3D representations to guide 3D optimization through diffusion models. These methods can guide outfit changes via images or text but are unable to model physics-related deformations.

Monocular reconstruction of layered NeRF and mesh representations of human avatars~\cite{Feng2023DELTA, Feng2022scarf} can experiment with clothing transfer, but cannot integrate physical simulation.
Some methods based on multi-layer 3D Gaussian Splatting representations~\cite{kerbl3Dgaussians}.~\cite{Huang2DGS2024, bian2024chatgarment} reconstruct multi-layered clothed human bodies, achieving high-fidelity clothing transfer effects. D3GA~\cite{zielonka25dega} and LayGA~\cite{lin2024layga} model the non-rigid deformation of clothing using pose- or time-dependent neural networks, while PhysAvatar~\cite{PhysAavatar24} and PICA~\cite{peng2024pica} embed physical simulations~\cite{grigorev2022hood, Li2021CIPC} to \textcolor{black}{generate garment animations with visually plausible cloth dynamics}. However, these methods all require multi-view sequences as input, which are not easily accessible to the general public.

\subsection{Consistent multi-view and video generation}

\textbf{Consistent multi-view generation.} 
Zero-1-to-3~\cite{liu2023zero1to3} pioneered camera-conditioned diffusion for zero-shot novel view synthesis. SyncDreamer~\cite{liu2023syncdreamer} and MVDream~\cite{shi2023mvdream} introduced 3D-aware synchronization and cross-view attention to enforce multi-view consistency. Wonder3D~\cite{long2023wonder3d} further improved geometric detail by jointly generating aligned normal and color maps. Current trends include discrete sequence modeling via ViewMask-1-to-3~\cite{zhu2025viewmask}  and autoregressive frameworks like MV-AR~\cite{hu2025mvar} that utilize preceding viewpoints as references to maintain consistency across widely separated views.

Human multi-view generation utilizes parametric models as geometric priors to handle non-rigid motion and complex clothing. Frameworks like PersonaCraft~\cite{kim2025personacraft} and Champ~\cite{zhu2024champ} employ 3D-aware control for occlusion-robust and shape-preserving animation. Advanced models such as D-Garment~\cite{dumoulin2026d} and Disco4D~\cite{Pang2025Disco4D} address dynamic clothing through physics-conditioned latent diffusion or disentangled Gaussian representations. Scaling to high resolutions, MEAT~\cite{wang2025meat} introduces efficient mesh attention for megapixel synthesis , while SyncHuman~\cite{chen2025synchuman} synchronizes 2D multi-view and 3D native branches to ensure structural integrity under challenging poses.

For multi-view generation tasks, generators trained on 2D data still suffer from inconsistencies across views, while those trained on 3D data are typically limited by the scarcity of high-quality 3D datasets. Therefore, we chose to leverage 2D generation methods to utilize vast amounts of video priors for 3D reconstruction.

\textbf{Video generation.}
Diffusion-based video generation has advanced rapidly in recent years. Stable Video Diffusion~\cite{blattmann2023stable}  leverages latent diffusion to significantly improve temporal consistency and resolution. Models such as OpenAI Sora~\cite{openai2024sora}, Google Veo 3~\cite{google2024veo3} and ByteDance Seedance~\cite{bytedance2024seedance} \textcolor{black}{generate long, high-resolution videos with high visual fidelity} using large-scale training and spatiotemporal Transformers~\cite{vaswani2017attention}, while Wan~\cite{wan2025wan} and HunyuanVideo~\cite{kong2024hunyuanvideo} promote open-source progress.
These models serve as general-purpose video generation foundations and also possess certain capabilities for human video generation.

Human video generation is evolving from 2D motion transfer toward physically-aware simulations with environment affordance and 3D geometric constraints. Animate Anyone~\cite{AnimateAnyone} established the technical foundation for identity consistency via ReferenceNet; MagicAnimate~\cite{MagicAnimate} improved detail fidelity through dense features and joint image-video training. Champ~\cite{zhu2024champ} introduced 3D parametric guidance (SMPL) to solve shape alignment and complex pose control. MimicMotion~\cite{mimicmotion2024} optimized motion stability and hand precision using confidence-aware guidance. Animate Anyone 2~\cite{hu2025animate} pioneered environment-coherent interactions. 
While human video generation is mostly concerned with how the person moves, it lacks the ability to swap or change outfits.

\section{Method}

% Our goal is to reconstruct a high-fidelity human avatar in terms of geometry, appearance, and physical properties from monocular videos, going beyond previous methods that primarily focused on geometry and appearance reconstruction.
% Reconstructing the physical properties of digital humans enhances realism while improving the model's generalization capabilities across different scenarios.
% To achieve the reconstruction of physical properties from a single camera, we build a high-fidelity personalized geometric template rather than SMPL~\cite{loper2015smpl} or SMPLX~\cite{SMPLX_2019}, and integrate physical simulation.
% However, geometry with only physical embedding cannot achieve high-fidelity rendering effects, as artifacts may appear in regions that are occluded or unseen in the input images when applied to new scenarios.
% To improve rendering generalization, we construct a key pose set to generate 2D data and combine it with real data for joint training.

\subsection{Algorithm Pipeline}

Given a monocular video (or a video generated from a single image) of a dynamically clothed human, denoted as $\{I_t | t = 1, \dots, N\}$, our goal is to reconstruct decoupled clothing and the underlying body which is ready for physical simulation and virtual try-on.
To achieve this, we reconstruct a clothed human represented by a double-layer Mesh-GS representation defined in Sec.~\ref{base representation} from the monocular inputs, and integrate the physical simulation in Sec.~\ref{sec: physical}.
% Since the unobserved regions in real data cannot be reconstructed directly, we generate data for joint training using a generative model with the key pose constructed from a large-scale motion dataset.
Since the unobserved regions in real data cannot be directly reconstructed, we generate data via generative models for joint training and address inter-frame inconsistency in generated data through geometric and appearance consistency constraints.

\begin{figure*}[ht]
  \centering
  \includegraphics[width=1.0\linewidth]{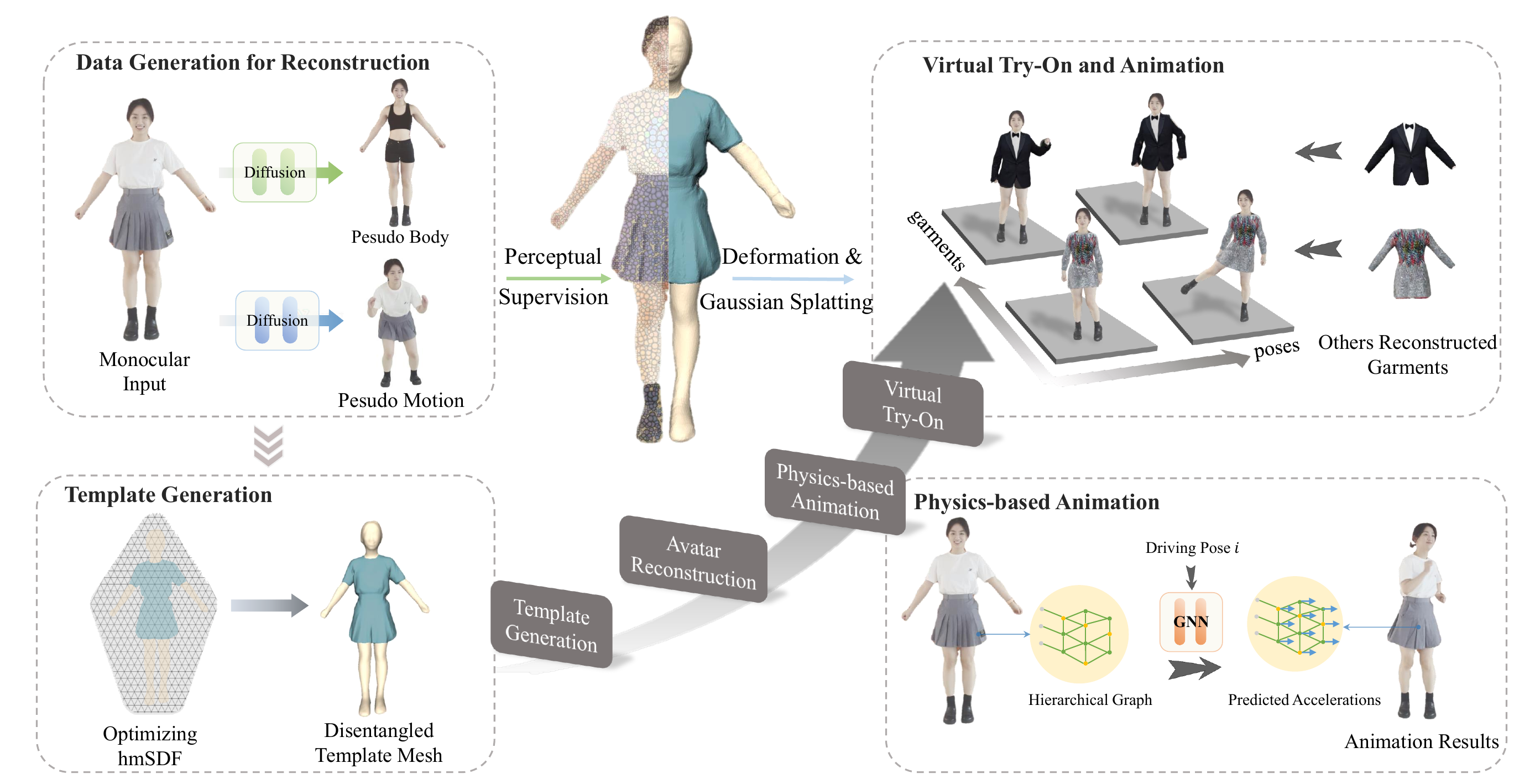}
  \begin{picture}(0,0)
    % \put(-175,-8){(a)}
    % \put(-60,-8){(b)}
  \end{picture}
  \vspace{-6mm}
  \caption{Overview of our approach.
  Our method can reconstruct double-layer Mesh-GS clothed human avatar from monocular inputs and apply it to \textcolor{black}{virtual try-on through physical simulation, producing visually plausible cloth dynamics}. We first construct a personalized geometric template and reconstruct high-fidelity geometry and appearance even for unseen area by jointly using generated data and ground-truth data. The reconstructed clothing and body can be used for novel pose animation and virtual try-on with the help of physical simulator.
    }
  \label{pipe}
\end{figure*}

% For unobserved regions, we construct a key pose set from a large-scale motion dataset and generated data for these poses using a generative model. 
The reconstructed double-layer clothed human avatar demonstrates strong generalization capabilities for novel clothing and poses, achieving high-fidelity virtual try-on, as we illustrated in our application in Sec.~\ref{apply}.
The overall pipeline of our approach is illustrated in Figure~\ref{pipe}.

\subsection{Disentangled Human Avatar Representation}
\subsubsection{Personalized Template Generation}
\label{sec:template rec}
We first reconstruct layered clothing and body meshes to provide a good initialization for 2DGS and integrating physical simulation.
Following D\(^3\)-Human~\cite{chen2025d3human}, we reconstruct the decoupled  body and clothing geometries \(X^{rec}\).
To remove pose and shape deformations, we first convert the star pose in the canonical space of D\(^3\)-Human to a T-pose and then eliminate shape deformations using the following formula:
\begin{equation}
    X^0 = \left(\sum_{j=1}^{J} w_{i,j}G_j(\theta)\right)^{-1} X(\theta, \beta) - \sum_{k=1}^{B} \beta_k \, \psi_k,
\end{equation}
% \[
% X^0 = \left(\sum_{j=1}^{J} w_{i,j} \, G_j(\theta)\right)^{-1} X(\theta, \beta) 
% - \sum_{k=1}^{B} \beta_k \, S_k,
% \]
where \(X(\theta, \beta)\)  represents the vertices of clothed human mesh \(X^{rec}\) under the pose \(\theta\) and shape \(\beta\), \(X^0\) is the template mesh without pose and shape deformations, \(w_{i,j}\) denotes the skinning weight of vertex \(i\) for joint \(j\), \(G_j(\theta)\) is the transformation matrix of joint \(j\) given the pose parameters \(\theta\), and \(\psi_k\) is the \(k\)-th shape blend shape obtained via nearest-neighbor interpolation. 
% Using \(X^0\), we obtain the decoupled body template \(G_b\) and clothing template \(G_c\).
Using \(X^0\), we obtain the decoupled body template and clothing template.

\subsubsection{Base Representation}
\label{base representation}

Similar to PICA~\cite{peng2024pica}, We anchor 2DGS on the faces of the template meshes to represent clothed avatars.
In PICA, the geometric properties of the GS (mean position, scaling, and rotation) are parameterized solely in terms of the mesh properties. 
Although this representation can support driving the reconstructed 2DGS clothed avatar with the simulator~\cite{grigorev2024contourcraft}, its capability is limited by the mesh resolution, resulting in low geometric detail.
To address this issue, we propose a global-local representation of geometric properties.
The global geometric properties ensure that the 2DGS moves along with the underlying template meshes, while the local geometric properties increase geometric detail.

\textbf{Global Properties.}
Given the position of the vertices ${\mathbf{p}}_{i j}$ of the $i$-th face ($j$=1,2,3), the global mean $\mathbf{x}_g$, the global rotation matrix $\mathbf{R}_g$ and the global scaling $\mathbf{S}_g$ of the 2DGS anchored to this face are defined in terms of their predefined barycentric coordinates $b_j$:
\begin{equation}
\begin{gathered}
\label{formula_global}
\mathbf{x}_g=\sum_{j=1}^3 b_j \mathbf{p}_{i j},\quad
\mathbf{R}_g = [\mathbf{r_1},\mathbf{r_2},\mathbf{r_3}],\\
\mathbf{r_1} = \mathbf{n_i},\quad
\mathbf{r_2}=\frac{\left(\mathbf{p}_{i1}-\mathbf{p}_{i0}\right)}
{\left\|\mathbf{p}_{i1}-\mathbf{p}_{i0}\right\|}, \quad 
\mathbf{r}_{3} = \mathbf{r}_{1} \times \mathbf{r}_{2},\\
\mathbf{S}_g = [0, \left\|\mathbf{p}_{i1}-\mathbf{p}_{i0}\right\|, \left\|\mathbf{p}_{i2}-\mathbf{p}_{i0}\right\|],
\end{gathered}
\end{equation}
where $\mathbf{n_i}$ is the normal of the face, [$\mathbf{r_1}$,$\mathbf{r_2}$,$\mathbf{r_3}$] refers to \textcolor{black}{the orthonormal bases} of the i-th face. 

\textbf{Local Properties.}
In addition to the global geometric properties, each 2DGS has the learnable local properties to increase the geometric detail. Formally, the final mean position $\mathbf{x}$, the rotation matrix $\mathbf{R}$ and the scaling $\mathbf{S}$ of the 2DGS is given by:
\begin{equation}
\begin{gathered}
\label{formula_local}
\mathbf{x}= \mathbf{x}_g + \mathbf{x}_l,\quad
\mathbf{R} = \mathbf{R}_g \cdot \mathbf{R}_l,\quad
\mathbf{S} = \mathbf{S}_g \cdot \mathbf{S}_l,
\end{gathered}
\end{equation}
% For more details, please refer to PICA~\cite{peng2024pica} for details.
where, $\mathbf{x}_l$, \textcolor{black}{$\mathbf{R}_l$ and $\mathbf{S}_l$} are the local position, rotation and scaling, respectively.

\textbf{Non-rigid Deformation.}
To better capture the per-frame details such as the wrinkle of the clothing when reconstructing the 2DGS and the underlying template meshes, we define a non-rigid deformation field \textcolor{black}{before transferring the avatar} from canonical T-pose to the observed pose by LBS.
The final transformation from the canonical space (T pose) to the current space can be formally written as:
\begin{equation}
\label{dfm}
\begin{gathered}
\hat{X}(\theta, \beta) = \sum_{j=1}^J w_{i,j} G_j(\theta) \cdot (X^0 +\Delta X^0 + \sum_{k=1}^{B} \beta_k \, \psi_k),\\
\Delta X^0 = f_{\textrm{dfm}} (X^0, e_t),
\end{gathered}
\end{equation}
% where \(\hat{X}(\theta, \beta)\) represents the mesh vertices in observed space with the SMPL-X~\cite{SMPLX_2019} pose parameter $\theta$ and shape parameter $\beta$. 
% \(X(\beta)\) is the corresponding T-pose mesh with the same shape \(\beta\), 
% \(w_{i,j}\) denotes the skinning weight of vertex \(i\) for joint \(j\), and \(G_j(\theta)\) is the transformation matrix of joint \(j\) given the pose parameters \(\theta\). 
where $\Delta X^0$ is the non-rigid deformation predicted by an MLP $f_{\textrm{dfm}}$, which is conditioned on the per-frame learnable vector $e_t$.
$X^0$ refers to the clothing and body meshes in canonical space, which is initialized in Sec~\ref{sec:template rec} and is optimized during training.
We use two MLPs to better model the non-rigid deformation of the body and clothing separately. 
Note that $X^0$ is shared across both real captured data and generated data during training. Conversely, the non-rigid deformation and shape-based deformation are independent for real and generated  data. This separation helps isolate the deformations of real and generated data, thereby addressing the geometric inconsistencies inherent in generated data.

\textbf{Rendering.}
Similar to PICA~\cite{peng2024pica}, we first render an albedo map $A_{\text{GS}}$ from the 2DGS:
\begin{equation}
\label{eq:render}
C=\sum_i\left(\alpha_i^{\prime} \prod_{j=1}^{i-1}\left(1-\alpha_j^{\prime}\right)\right) c_i,
\end{equation}
where $C$ is the rendered pixel color, $\alpha_i^{\prime}$ is the weighted opacity according to the probability density of the i-th projected GS at the target pixel. 
$c_i$ is the color computed from the SH coefficients.
A normal map $N_\text{GS}$ is \textcolor{black}{rendered together with the albedo map}.
Unlike PICA~\cite{peng2024pica} that uses a U-Net~\cite{ronneberger2015u} to model the view-dependent self-shading from the normal map, we directly calculate the shading map through spherical harmonics (SH)~\cite{green2003spherical, ramamoorthi2001efficient}.
This is because learning view- and pose-dependent shading \textcolor{black}{without appropriate priors remains challenging} due to monocular inputs.
The final image is given by:
\begin{equation}
    \label{eq:shading}
    I = A_{\text{GS}} \cdot I_{\text{shading}},\quad I_{\text{shading}} = SH(N_\text{GS}, sh),
\end{equation}
Where $SH(\cdot)$ refers to the SH evaluation, and $sh$ is a learnable SH coefficient representing the lighting condition.

\subsection{Key Pose Extraction}
\label{sec:key pose}
We aim to use the generated human image sequences as supplementary data to enhance the performance of reconstructed clothed humans in novel poses and garments. However, human poses are highly diverse, and due to the computational cost of generation and training, it is infeasible to directly use large-scale pose data to generate sequences and train the model.

To ensure that garments are fully observed and optimized under various typical body configurations, we design a set of key poses aimed at maximizing the coverage of the human motion space (e.g., turning, arm-raising, and walking). The specific variations introduced by different garments—such as unique wrinkles, folding patterns, and physical behaviors—are explicitly modeled by our double-layer disentangled representation, physics simulation module, and deformation representation. This approach ensures the universality and robustness of the key pose set across diverse garment categories.

We selected the key pose set by considering the following three metrics: \textit{Pose Representativeness}, \textit{Pose Similarity}, and \textit{Rendering Visibility}. By combining these three indicators, we aim to ensure that the selected poses not only capture the essential variation of human motion but also exhibit sufficient diversity and possess informative visual appearance.

\textbf{Pose Representativeness.} We employ Kernel Density Estimation (KDE) on the dimension-reduced pose data to estimate the probability density $p(\theta)$, for each pose \(\theta\). A higher \(p(\theta)\) indicates that the pose is more representative in the dataset.

\textbf{Pose Similarity.}
To avoid selecting too many similar poses and thus increase the overall diversity, we use the cosine similarity. For two poses \(\theta_i\) and \(\theta_j\), the cosine similarity is given by
\[
\text{sim}(\theta_i,\theta_j) 
= \frac{\theta_i \cdot \theta_j}{\|\theta_i\|\;\|\theta_j\|}.
\]

\textbf{Rendering Visibility.}
Since we want the generated data to provide as much information as possible about the human avatar, we use SMPL-X~\cite{SMPLX_2019} as a proxy, and render each vertex's visibility for a given pose and sum these visibilities to obtain a visibility score:
\[
v(\theta) = \sum_{k=1}^{K} \mathbb{I}_{\text{visible}}(\text{vertex}_k|\theta),
\]
where \(K\) is the total number of SMPL-X vertices and \(\mathbb{I}_{\text{visible}}(\cdot)\) indicates whether a vertex is visible in the rendered pose.

\textbf{Pose Extraction via MMR.}
In order to comprehensively consider the representativeness and visibility of candidate poses, as well as their similarity to the selected set, we employ the Maximal Marginal Relevance (MMR)~\cite{carbonell1998use} principle to define a comprehensive scoring function.
% Specifically, let \(p(\theta)\) denote the pose's probability density (representativeness), \(v(\theta)\) the pose's rendering visibility score, and \(\mathrm{sim}\bigl(\theta,\theta_j\bigr)\) the similarity (or distance) measure between poses \(\theta\) and \(\theta_j\). 
Given a set of already selected poses \(\mathcal{E}\), the MMR-based comprehensive score for a candidate pose \(\theta_i\) is formulated as
\begin{equation}
    \mathrm{Score}(\theta_i)
    \;=\;
    \alpha\, p(\theta_i)
    \;+\;
    \beta\, v(\theta_i)
    \;-\;
    \gamma \,\max_{\theta_j \in \mathcal{E}}\,
    \mathrm{sim}\!\bigl(\theta_i,\theta_j\bigr),
\end{equation}

Here, \(\alpha,\;\beta,\;\gamma\) are weight parameters.
At each step, we select the pose \(i \notin \mathcal{E}\) that maximizes
% In practice, we adopt a greedy algorithm at each selection step by choosing the pose \(i \notin S\) that maximizes
\[
\max_{i \notin \mathcal{E}}
\Bigl(
    \alpha\, p(\theta_i)
    \;+\;
    \beta\, v(\theta_i)
    \;-\;
    \gamma \,\max_{\theta_j \in \mathcal{E}}\,
    \mathrm{sim}\!\bigl(\theta_i,\theta_j\bigr),
\Bigr),
\]
% where \(S\) is the set of already selected poses, \(p_i\) denotes the probability density (representativeness) of pose \(i\), \(v_i\) is its rendering visibility score, and \(\mathrm{Dist}(i,j)\) measures the distance (or dissimilarity) between poses \(i\) and \(j\). 
We select 500 key poses from 3.45 million poses in the CMU subset of AMASS~\cite{AMASS:ICCV:2019} using the MMR principle to generate training data.
This selection process ensures that the final key pose set achieves a good balance among representativeness, visibility, and diversity.

\subsection{Data Generation}
\label{sec:data gen}
\textbf{Key Pose Data Generation.}
% When driving clothing, there are not only occluded regions (e.g., underarms) but also wrinkles and deformations caused by motion. 
% Therefore, we construct more diverse generated data for the key pose data.
% Using the extracted key poses, we generate clothed human image sequences for specified pose sequences through a human animation generation method~\cite{mimicmotion2024}.
% Due to the significant differences between the extracted poses, there are discontinuities between frames, which can affect the quality of the generated image sequences.
% Therefore, we perform interpolation between frames to generate a continuous image sequence.
When driving the clothed human body to new motions, there may be large invisible regions (for single image input) or occluded regions (such as underarms, for monocular video input).
% These regions are invisible in the real captured data, which can lead to artifacts when animating the reconstructed human avatar to novel poses.
To address this, we construct more diverse generated data for the key pose data.
Using the extracted key poses, we generate clothed human image sequences for specified pose sequences through a human animation generation method~\cite{viggle2025}.
Since the generated data includes diverse poses, it effectively provides information for the unseen regions, helping to eliminate artifacts.

\textbf{Body Data Generation.}
Since we reconstruct decoupled clothing and the human body, rendering artifacts may appear in unobserved body regions when applying novel clothing or novel poses. For example, when changing a long-sleeved outfit to a short-sleeved one, artifacts may occur on the exposed arms, as well as on body parts revealed by clothing movement during motion.

\textcolor{black}{To plausibly complete underlying-body regions that are occluded by clothing or otherwise unobserved in the monocular input, we first employ a diffusion-based image inpainting method to redraw the clothed person as a minimally clothed person}.
Then we use a 3D human generation method~\cite{viggle2025} to create 360-degree generated data for the minimally clothed body and estimate SMPL-X parameters~\cite{moon2024exavatar} for the redrawn body generated data.

\subsection{Generated Data Alignment.}
% Since the generated image sequence does not fully align with the input poses, we perform re-tracking~\cite{moon2024exavatar} on the generated image sequence to obtain accurate body poses.
% In addition, there is a difference between the generated data and the real capture data in both geometry detail and appearance style. 
% To fill this difference, we use another non-rigid deformation MLP defined in Eq.~\ref{dfm} to adaptively learn the correspondence between the model and the generated data, 
% and use only the perceptual term~\cite{zhang2018unreasonable} instead of the full photometric loss for the generated data.    
Due to the inevitable inter-frame inconsistency in generated image sequences, as well as the discrepancies between generated and real data, directly using such data for human avatar reconstruction introduces significant artifacts. 
To address this, we decompose the inter-frame inconsistency in generated sequences into geometric inconsistency and appearance inconsistency.
Specifically, geometric inconsistency comprises shape discrepancy and human pose misalignment, while appearance inconsistency manifests as subtle style variations and shading incoherence.

\textbf{Shape discrepancy.}
When generating human animation sequences, since 2D generation methods cannot guarantee accurate consistency—including both consistency with real data and inter-frame consistency of the generated data—two main issues arise: the body shape in the generated data cannot perfectly match the real data, and the corresponding 3D human geometry may vary between frames, manifesting as inconsistent garment lengths. To address the discrepancy between generated and real data, we re-optimize the shape parameters for the generated data. For the inter-frame inconsistency, we employ an additional non-rigid deformation MLP defined in Eq.~\ref{dfm} to adaptively learn the correspondence between the model and the generated data.

\textbf{Human pose misalignment.}
Since the generated image sequence does not fully align with the input poses, we perform re-tracking~\cite{moon2024exavatar} on the generated image sequence to obtain accurate body poses.

\textbf{Shading incoherence.}
The appearance of the generated data lacks temporal consistency, especially when the generated background changes or shakes. 
This leads to unstable lighting and shading effects.
To solve this problem, we optimized the SH coefficients for each frame of the generated data to model the unstable lighting.
We only optimize the global SH coefficients, thus avoiding over-fitting to individual frames.
% While our reconstructed model only have one 

\textbf{Subtle style variations.}
Since generated image sequences may exhibit subtle stylistic variations (e.g., color drifting) and lack local feature coherence, leading to inaccurate 3D alignment and potential failure of per-pixel loss functions during reconstruction, we employ perceptual loss terms~\cite{zhang2018unreasonable} as supervision for the generated data to mitigate reconstruction errors caused by style discrepancies.

\subsection{Training}
% \subsubsection{Loss}
We first generate initial template meshes through D\(^3\)-Human.
Then the template meshes, the properties of the 2D Gaussians (local geometric properties, opacity and color), the deformation MLPs, and the SH coefficient are optimized.
% During training, we only use the input video as training data in the first 200 epochs, and then adding the generated data to refine the occluded regions.
The total loss can be \textcolor{black}{written} as:
% \begin{equation}
%     L=L_{\text{color}} +  L_{\text{label}} +  L_{\text{geo}} +  L_{\text{normal}} .
% \end{equation}
\begin{equation}
\begin{aligned}
\text{L} ={}&
\text{L}_{\text{color}}
+ \lambda_{\text{label}}\text{L}_{\text{label}}
+ \lambda_{\text{geo}}\text{L}_{\text{geo}} \\
&+ \lambda_{\text{normal}}\text{L}_{\text{normal}}
+ \lambda_{\text{reg}}\text{L}_{\text{reg}}
+ \textcolor{black}{\lambda_{\text{smooth}}\text{L}_{\text{smooth}}}.
\end{aligned}
\end{equation}
Here, the appearance loss $\text{L}_{\text{color}}$ \textcolor{black}{consists} of $\text{L}_{1}$ term, SSIM term, and perceptual term~\cite{zhang2018unreasonable}. 
For the real captured data, we use the full appearance loss, while for the generated data, we only use the perceptual loss for the generated data since the color style of the generated data is \textcolor{black}{slightly different from} the ground truth data, as we discussed above. 
$\text{L}_{\text{label}}$ and $\text{L}_{\text{geo}}$ are the label loss and the geometry loss from PICA~\cite{peng2024pica}, which are used to ensure each object (body,  clothing) is represented by the 2DGS positioned on the correct template meshes and a smooth geometry, respectively.

Inspired by D\(^3\)-Human~\cite{chen2025d3human}, we use the perceptual normal loss to further enhance the rendered normals:
\begin{equation}
\begin{aligned}
\text{L}_{\text{normal}} = \sum_i \|\phi_i(N_\text{GS}) - \phi_i(\hat{N})\|^2,
\end{aligned}
\end{equation}
where \(\hat{N}\) is the ground truth normal obtained through Sapiens~\cite{khirodkar2024sapiens}, and $\phi_i(*)$ denotes the activation of the \(i\)-th layer in the MobileNetV2 network ~\cite{sandler2018mobilenetv2}. 

$\text{L}_{\text{reg}}$ is a regularization loss which is used to encourage the anchored 2DGS aligned with the mesh faces:
\begin{equation}
\begin{aligned}
\text{L}_{\text{reg}} = \sum_{N_{GS}} (\| Q(\mathbf{R}_l) - [1,0,0,0]\|^2) + \| \mathbf{x}_l\|^2 + \| \mathbf{S}\|^2,
\end{aligned}
\end{equation}
% where $Q(\mathbf{R}_l)$ and $\delta$ refer to the quaternionic forms of the local rotation. 
\textcolor{black}{where $Q(\mathbf{R}_l)$ denotes the quaternion representation of the local rotation $\mathbf{R}_l$.}
For the generated data, there may be inconsistencies across frames that lead to high-frequency noise. 
We filter out this high-frequency information using a self-supervised smoothing loss term $\text{L}_{\text{smooth}}$,
\begin{equation}
\begin{aligned}
\text{L}_{\text{smooth}} = \sum_{i} \sum_{j \in N(i)} (\|c_i-c_j \|^2) + \| \alpha_i-\alpha_j \|^2),
\end{aligned}
\end{equation}
where $c_i$ and $\alpha_i$ are the color and opacity of the i-th 2DGS, $N(i)$ are the K-nearest neighbors of the i-th 2DGS.
\subsection{Physical Simulation}
\label{sec: physical}
After reconstructing the clothed avatar, \textcolor{black}{we use the pre-trained ContourCraft model~\cite{grigorev2024contourcraft} to generate animation results with visually plausible cloth dynamics through physical simulation}.

Formally, ContourCraft \textcolor{black}{predicts} per-vertex acceleration frame by frame with a GNN trained with self-supervised loss:
\begin{equation}
\label{eq:gnn}
    \begin{gathered}
    \Ddot{\mathbf{p}}^t = G(\mathbf{p}^{t-1} ,\dot{\mathbf{p}}^{t-1}, \mathbf{b}^{t-1}, \mathbf{b}^{t}, \mathbf{\rho}; \phi),\\
    \dot{\mathbf{p}}^{t} = \dot{\mathbf{p}}^{t-1} + \Ddot{\mathbf{p}}^t,\quad
    \mathbf{p}^{t} = \mathbf{p}^{t-1} + \dot{\mathbf{p}}^{t},
    \end{gathered}
\end{equation}
where $\Ddot{\mathbf{p}}$, $\dot{\mathbf{p}}$ and $\mathbf{p}$ are the acceleration, velocity and position of the vertices of the garment meshes, respectively. 
$\mathbf{b}$ is the position of the vertices of the body mesh.
$\mathbf{\rho}$ refers to the per-vertex material coefficient (mass, bending coefficient and Lame coefficient) of the garment and $G$ is the GNN with the learnable weight $\phi$.
Please refer to ContourCraft~\cite{grigorev2024contourcraft} for more details.

To fit the physical properties of the garment to the real captured input, we then optimize the material coefficient of the garment $\mathbf{\rho}$ and the learnable weight $\phi$ according to the reconstructed results:
\begin{equation}
\begin{gathered}
\mathbf{\rho}, \phi =\mathop{\arg\min}_{\mathbf{\rho}, \phi} \ \ \sum_{t=1}^{T}\left\|\Ddot{\mathbf{p}}^{t+1}_{r}-G(\mathbf{p}^{t}_{r},\dot{\mathbf{p}}^{t}_{r} , \mathbf{\rho}; \phi)\right\|_{2}^{2},\\
\end{gathered}
\end{equation}
Where $\mathbf{p}^{t}_{r}$ is the reconstructed position of the vertices of the clothing mesh at frame $t$, $\Ddot{\mathbf{p}^{t}_{r}}$ and $\dot{\mathbf{p}^{t}_{r}}$ refer to the acceleration and velocity, respectively.
\textcolor{black}{During the physics-fitting stage, we freeze the reconstructed avatar's geometry and appearance parameters, and optimize only the garment material parameters $\mathbf{\rho}$ and the learnable simulator parameters $\phi$.}

Once we have processed the garment meshes using the simulation process, all 2D Gaussians are transferred from the canonical space to the current space according to the formulas defined in Sec.~\ref{base representation}, and generate the novel pose driving and novel view synthesis results through Eq.~\ref{eq:shading}. 
The 2D Gaussians would be spiky if the surface is spiky while driving, since the scaling parameters of the 2D Gaussians depend on the local mesh surfaces.
To avoid this, we set the maximum value of the Gaussian scale to be no greater than $\gamma$ times the minimum value, where $\gamma$ is the hyperparameter, which we set to 3 in practice.
% \vspace{-3mm}
\begin{figure*}[htbp]
  \centering
  \includegraphics[scale=0.055]{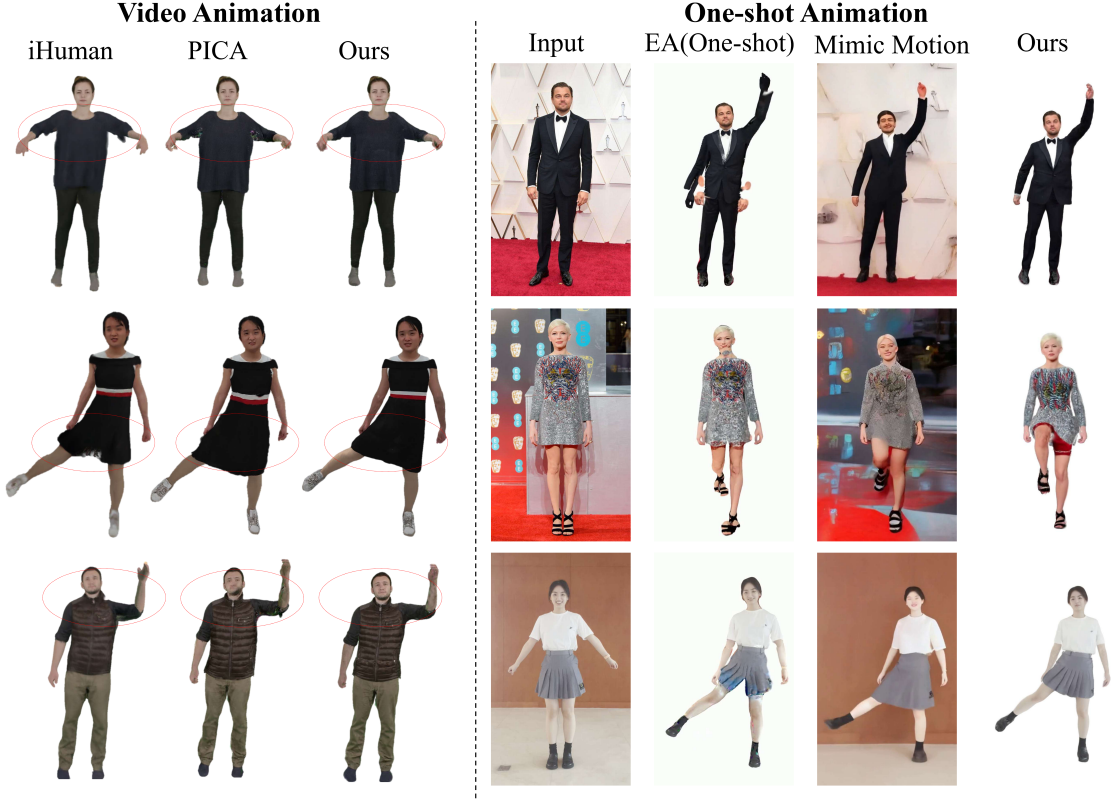}
  \begin{picture}(0,0)
    % \put(-175,-8){(a)}
    % \put(-60,-8){(b)}
  \end{picture}
  \vspace{-4mm}
  \caption{
  Qualitative comparison of novel pose animation. The left side uses a monocular video input, while the right side uses only a single image (one shot).
  EA (one-shot) represents the reconstruction result obtained by ExAvatar using a single image.
    }
  \label{comp_animation}
\end{figure*}

\vspace{-3mm}
\begin{figure*}[htbp]
  % \begin{minipage}{0.8\textwidth}  % 控制图片宽度
    % \hspace{0pt}  % 通过此方式手动设置水平位置
    \includegraphics[width=\linewidth]{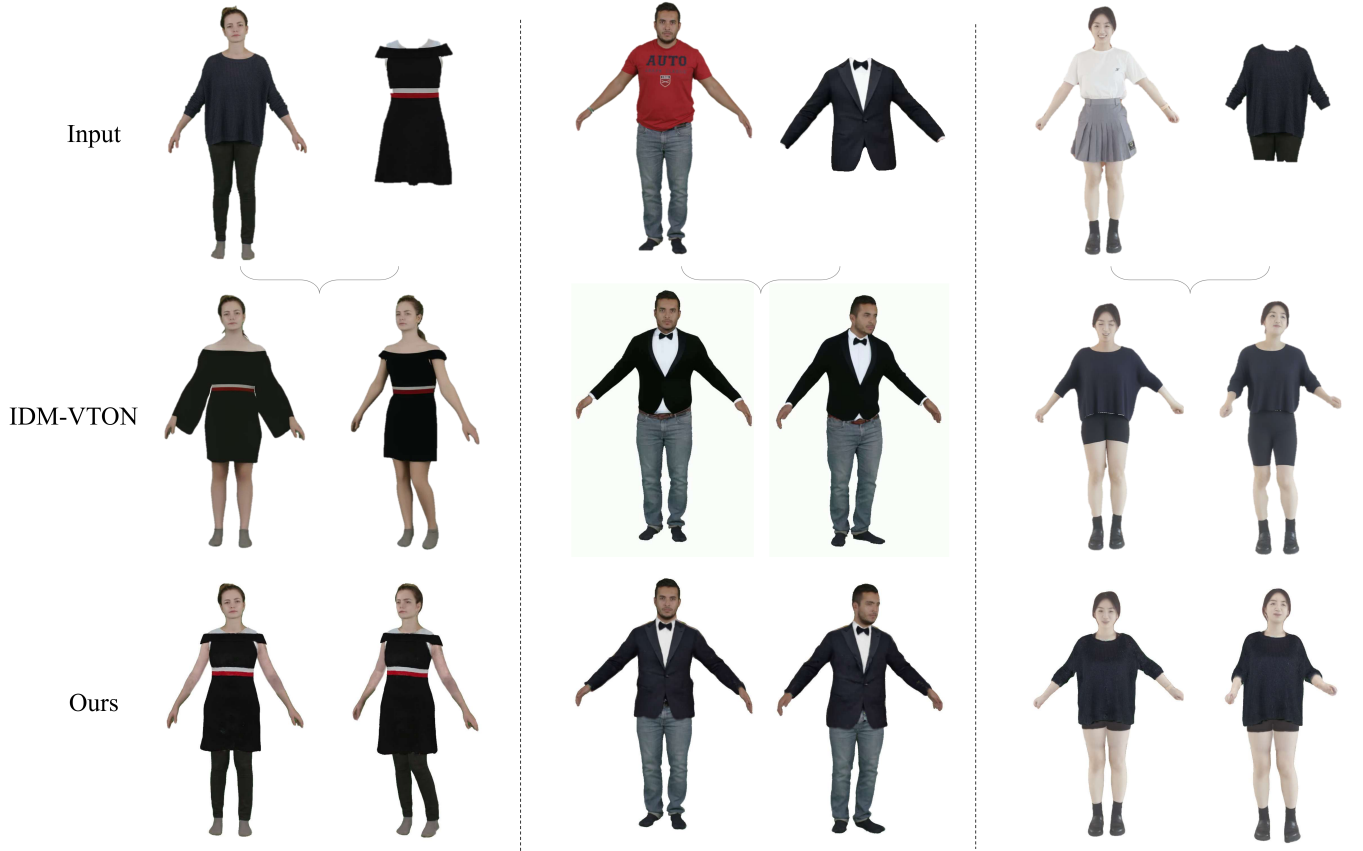}
  % \end{minipage}
  % \begin{picture}(0,0)
  %   \put(-445,98){Input}
  %   \put(-455,-8){IDM-VTON}
  %   \put(-445,-108){Ours}
  % \end{picture}
  \vspace{-23mm}
  \caption{
    Results of Virtual try-on. 
    We compare the try-on results with IDM-VTON using the same body, clothing, and pose.
    }
  \label{tryon}
\end{figure*}

\section{Experiments}
We conduct qualitative experiments for novel pose and quantitative experiments for reconstruction to demonstrate the effectiveness of our method. 
For the quantitative experiments, we use subjects from public dataset wearing different clothes to evaluate the quality of reconstruction. 
For qualitative experiments, we demonstrated the performance of driving reconstructed human avatars to novel motions. Additionally, we conducted ablation studies to showcase the effectiveness of keyframe extraction, generated data and learnable local properties. For applications, we presented the results of virtual try-on and relighting.
% The dataset and some experimental results are presented in the Supplementary Material.
More dynamic results can be found in the video.

\subsection{Implementation Details}
Our framework is implemented in PyTorch based on the 3DGS infrastructure, where we follow PICA~\cite{peng2024pica} and anchor 13 Gaussians on each triangular face of the reconstructed meshes. 
For data preprocessing of monocular video or single-image inputs, we predict the initial SMPL-X parameters via ExAvatar~\cite{moon2024exavatar} and extract semantic segmentation masks using SAM2~\cite{ravi2024sam2} and Sapiens~\cite{khirodkar2024sapiens}, while employing 2D generative priors (e.g., Viggle AI) to generate synthesized data sequences. The optimization of our joint Mesh-GS representation consists of two sequential stages; first, the deformation and appearance compensation networks are trained alongside the Gaussian attributes to eliminate generative inconsistencies, which takes 100K iterations to converge and requires about 10 hours on a single NVIDIA RTX 3090 GPU for a standard sequence of 200 real frames and 200 MMR-extracted frames. 
Second, for video inputs, the physical property fitting stage (Eq. 13) refines the cloth simulator parameters by optimizing over the training motion sequence, taking an additional 1 \textcolor{black}{hour}, whereas for single-image inputs, this physics fitting stage is bypassed ($T=1$) to directly adopt the default physical parameters of the pre-trained model. 
For downstream inference and simulation during novel pose animation and virtual try-on applications, we animate the underlying human body using Linear Blend Skinning (LBS), while the decoupled garment mesh dynamics are concurrently simulated using the fine-tuned physical parameters within a pre-trained GNN-based cloth simulator, achieving a total rendering speed of approximately 4 frames per second (fps) on a single NVIDIA RTX 3090 GPU.

\subsection{Dataset}
Since there are no publicly available real datasets for evaluating the quality of virtual try-on from monocular videos, we use subjects provided by PeopleSnapshot~\cite{alldieck2018video} and SelfRecon~\cite{jiang2022selfrecon} 
and REC-MV~\cite{qiu2023RECMV}. 
For monocular video inputs, we quantitatively compare the performance of our method and GS-based human reconstruction methods iHuman~\cite{pramishp2024iHuman} and PICA~\cite{peng2024pica}.
For single image input, \textcolor{black}{we compare our method with} diffusion-based human generation method MimicMotion~\cite{mimicmotion2024}, and the mesh-based reconstruction method ExAvatar~\cite{moon2024exavatar}.
Specifically, we perform our experiment on three subjects (female-3-casual, male-2-casual, male-3-casual) of People Snapshot dataset~\cite{alldieck2018video} and two subjects (CHH-female, anran-dance) wearing skirts from SelfRecon~\cite{jiang2022selfrecon} and RecMV~\cite{qiu2023RECMV}, respectively.
For each subject, we predict the SMPL-X parameters through \textcolor{black}{ExAvatar}~\cite{moon2024exavatar}, and generate the semantic segmentation maps with the help of SAM2~\cite{ravi2024sam2} and Sapiens~\cite{khirodkar2024sapiens}. 
We use the official codes of the baseline method.
For a fair comparison, we use the SMPL parameters transformed from the SMPL-X parameters predicted by \textcolor{black}{ExAvatar} for iHuman~\cite{pramishp2024iHuman}.

\begin{figure*}[htbp]
  \centering
  \vspace{-1mm}
  \includegraphics[width=\linewidth]{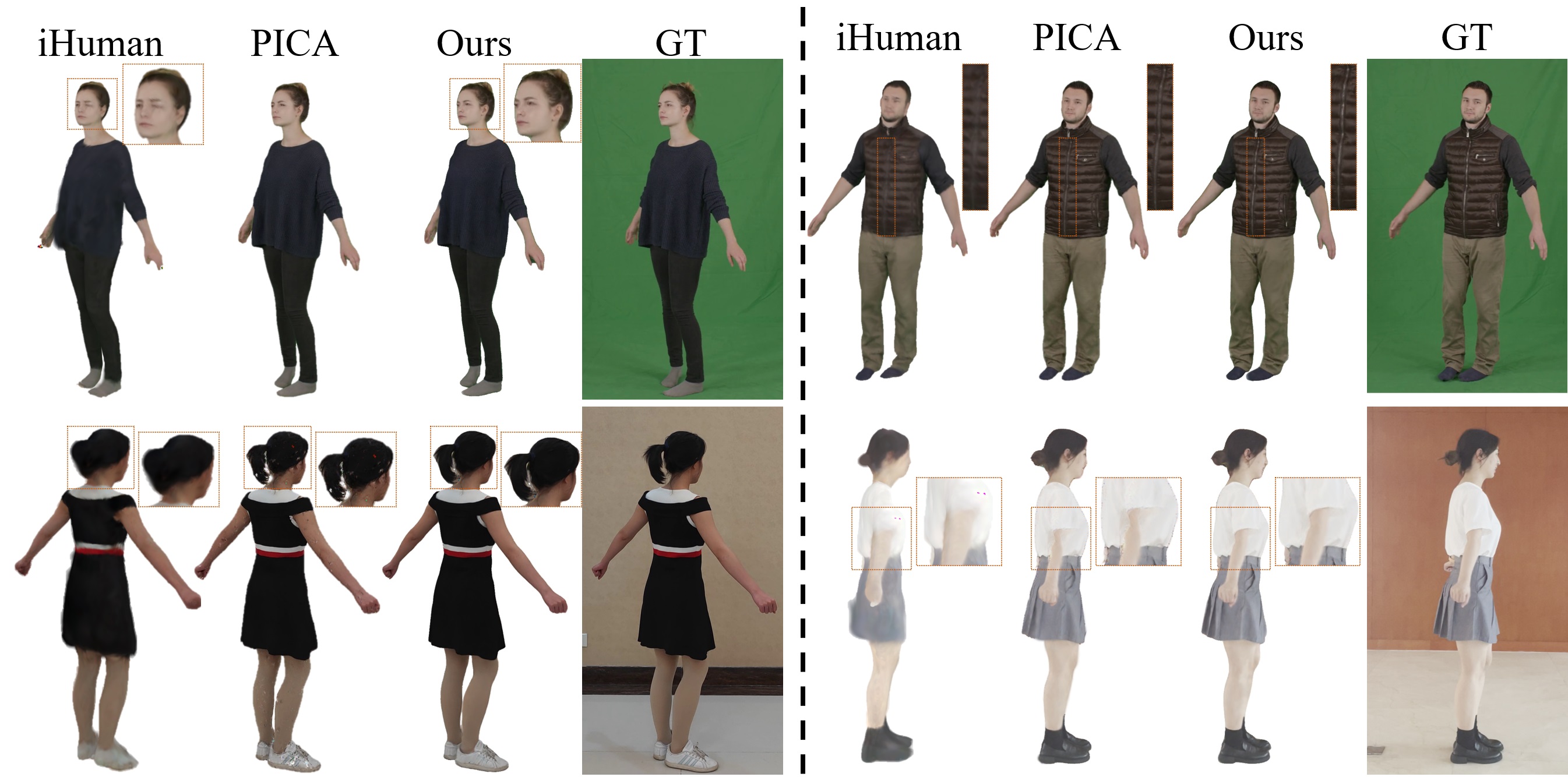}
  \begin{picture}(0,0)
    % \put(-175,-8){(a)}
    % \put(-60,-8){(b)}
  \end{picture}
  \vspace{-2mm}
  \caption{
  Qualitative comparison of monocular video reconstruction.
    }
  \label{comp_recon}
  % \vspace{-12mm}
\end{figure*}

% \vspace{-3mm}
\begin{table}[h]
\centering
\caption{Quantitative comparison with PICA and iHuman.}
% \vspace{-3mm}
\label{tab1}
\begin{tabular}{ccccc}
\hline
Subject              & Metric & iHuman      & PICA        & Ours        \\
\hline
\multirow{3}{*}{m2c} & PSNR/dB$\uparrow$    & 24.98 & 29.61 & \textbf{32.87} \\
                     & SSIM$\uparrow$    & 0.9476 & 0.9637  & \textbf{0.9850} \\
                     & LPIPS$\downarrow$   & 0.0734 & 0.0498 & \textbf{0.0231}  \\
 \hline
\multirow{3}{*}{m3c} & PSNR/dB$\uparrow$    & 24.70 & 29.45 & \textbf{30.98}  \\
                     & SSIM$\uparrow$    & 0.9621 & 0.9746 & \textbf{0.9830}  \\
                     & LPIPS$\downarrow$   & 0.0598 & 0.0232 & \textbf{0.0230} \\
\hline                    
\multirow{3}{*}{f3c} & PSNR/dB$\uparrow$    & 24.81 & 30.92 & \textbf{32.49} \\
                     & SSIM$\uparrow$    & 0.9664 & 0.9795 & \textbf{0.9876} \\
                     & LPIPS$\downarrow$   & 0.0648 & 0.0265 & \textbf{0.0198} \\
\hline
\multirow{3}{*}{chh} & PSNR/dB$\uparrow$    & 25.12 & 28.03 & \textbf{31.91} \\
                     & SSIM$\uparrow$    & 0.9720  & 0.9779 & \textbf{0.9859} \\
                     & LPIPS$\downarrow$   & 0.0435  & 0.0216 & \textbf{0.0205} \\ 
\hline
\multirow{3}{*}{anran} & PSNR/dB$\uparrow$    & 26.19 & 30.23 & \textbf{33.01} \\
                     & SSIM$\uparrow$    & 0.9579  & 0.9718 & \textbf{0.9878} \\
                     & LPIPS$\downarrow$   &  0.0771  & 0.0323 & \textbf{0.0224} \\
\hline              
\end{tabular}
\end{table}

% \vspace{-3mm}
\begin{table}[h]
\centering
\caption{Quantitative comparison with MimicMotion and ExAvatar.}
\vspace{-3mm}
\label{tab2}
\begin{tabular}{ccccc}
\hline
Subject              & Metric & \makecell{Mimic-\\Motion}    & \makecell{ExAvatar \\ 1-shot}       & Ours        \\
\hline
\multirow{3}{*}{m2c} & PSNR/dB$\uparrow$    & 24.17 & 21.44 & \textbf{26.02} \\
                     & SSIM$\uparrow$    & 0.9307 & 0.9139  & \textbf{0.9417} \\
                     & LPIPS$\downarrow$   & 0.0701 & 0.0837 & \textbf{0.0590}  \\
 \hline
\multirow{3}{*}{m3c} & PSNR/dB$\uparrow$    & 24.63 & 20.89 & \textbf{26.57}  \\
                     & SSIM$\uparrow$    & 0.9394 & 0.9233 & \textbf{0.9575}  \\
                     & LPIPS$\downarrow$   & 0.0637 & 0.0810 & \textbf{0.0465} \\
\hline                    
\multirow{3}{*}{f3c} & PSNR/dB$\uparrow$    & 25.48 & 21.58 & \textbf{28.30} \\
                     & SSIM$\uparrow$    & 0.9599 & 0.9467 & \textbf{0.9722} \\
                     & LPIPS$\downarrow$   & 0.0628 & 0.0702 & \textbf{0.0417} \\
\hline
\multirow{3}{*}{chh} & PSNR/dB$\uparrow$    & 24.55 & 19.94 & \textbf{25.32} \\
                     & SSIM$\uparrow$    & 0.9611  & 0.9493 & \textbf{0.9755} \\
                     & LPIPS$\downarrow$   & 0.0400 & 0.0606 & \textbf{0.0346} \\ 
\hline
\multirow{3}{*}{anran} & PSNR/dB$\uparrow$  & 26.76   & 23.75 & \textbf{27.84} \\
                     & SSIM$\uparrow$    &  0.9593 & 0.9451 & \textbf{0.9694} \\
                     & LPIPS$\downarrow$   &  0.0548  & 0.0739 & \textbf{0.0493} \\
\hline              
\end{tabular}
\end{table}

\subsection{Quantitative Evaluation}
We perform a quantitative experiment to evaluate our reconstruction quality.
Following PICA~\cite{peng2024pica}, we use three metrics to quantify the reconstruction results: structural similarity index (SSIM), peak signal-to-noise ratio (PSNR) and the learned perceptual image patch similarity (LPIPS)~\cite{zhang2018unreasonable}. 

\textbf{Reconstructing from monocular videos.}
Since there is no ground truth mesh or images from another view, we extract a subset of frames for each sequence for training and use the remaining frames for testing.
iHuman~\cite{pramishp2024iHuman} uses monocular sequences to reconstruct non-disentangled clothed humans.
PICA uses multi-view sequences to reconstruct clothing and garments, and we adapt it to monocular sequences.
\textcolor{black}{For the monocular PICA baseline, we provide only one view for each frame instead of the original multi-view observations, while keeping the original PICA representation and optimization scheme unchanged. We use the same preprocessing and body-pose initialization as our method.}
Our method achieves the best performance on all metrics, as shown in Table~\ref{tab1}.

\textbf{Reconstructing from single images.}
For single image inputs, we select a frame from the provided real sequence data, construct generated data for reconstruction, and evaluate on the real sequence.
MimicMotion~\cite{mimicmotion2024} uses a single image as input to generate image sequences of specified poses. However, the generated results exhibit poor consistency in body shape and style. 
ExAvatar~\cite{moon2024exavatar} is based on the SMPL-X representation and reconstructs from a single image in this experiment, while it fails to accurately restore textures in regions unseen in the input image, which highlights the importance of generating information for unseen regions.
Our method, with the assistance of the inconsistency elimination strategy, can reconstruct high-fidelity human avatars from single images and achieves superior performance across all metrics,  as shown in Table~\ref{tab2}.

% \begin{figure*}[ht]
%   \centering
%   \includegraphics[scale=0.32]{Figs/compsv2.pdf}
%   \begin{picture}(0,0)
%     % \put(-175,-8){(a)}
%     % \put(-60,-8){(b)}
%   \end{picture}
%   \vspace{-2mm}
%   \caption{
%   Compare the results of reconstruction (left) on testing frames and novel pose animation (right).
%     }
%   \label{comp1}
% \end{figure*}

% \vspace{3mm}
\subsection{Qualitative Evaluation}
For monocular video \textcolor{black}{inputs}, We present a comparison of the reconstructed results of the test frames of Figure~\ref{comp_recon}. The rendering results of iHuman are relatively blurry. Our rendering results have similar clarity to PICA but with fewer artifacts.
% In the qualitative experiments, 
We compare animation results of motion sequences that differ significantly from the input video. 
This demonstrates the generalization ability of the human avatars reconstructed by our method to novel poses.
We present the animation results of the novel poses of Figure~\ref{comp_animation}.
iHuman relies on LBS for animation, which does not produce natural animations for loose clothing. 
For example, artifacts may appear in the skirt when the legs are spread apart, as the actual skirt does not follow the skeletal LBS deformation.
PICA reconstructs a double-layer human body without using generated data, potentially causing artifacts in unseen areas.
MimicMotion can directly generate animation sequences corresponding to given poses, but cannot effectively maintain body shape and image style.
ExAvatar completely lacks texture information when exposing unseen regions.
With the help of physical simulation and generated data, our method generates natural animations for new poses without introducing artifacts.

\begin{figure}[htbp]
  \centering
  \includegraphics[scale=0.4]{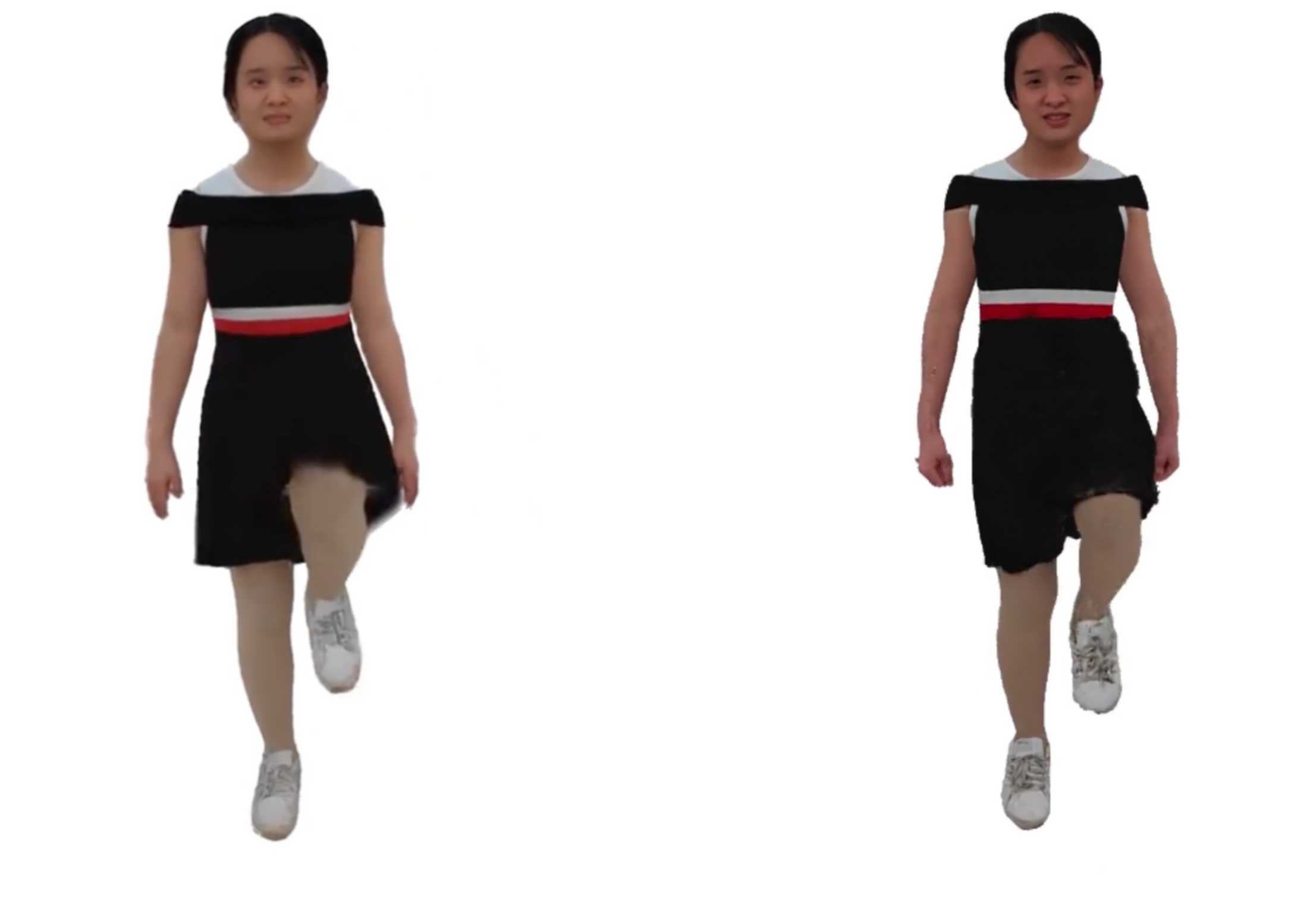}
  % \vspace{-5mm}
    \begin{picture}(0,0)
    \put(-160,-1){LHM}
    \put(-50,1){Ours}
  \end{picture}
  \vspace{-1mm}
  \caption{Qualitative comparison with LHM on novel pose animation.}
  \label{fig:lhm_comparison}
\end{figure}
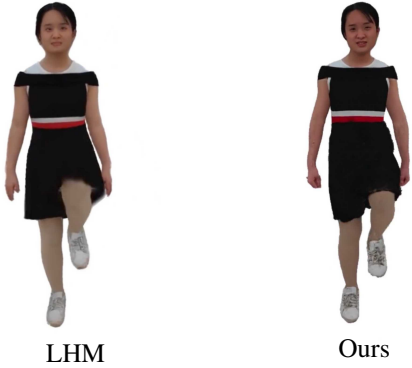

Furthermore, we compare our method with the recent state-of-the-art generalizable human reconstruction model LHM~\cite{qiu2025LHM}. While LHM can reconstruct detailed human avatars efficiently, its underlying animation heavily relies on conventional skeletal Linear Blend Skinning (LBS) without physical priors. Consequently, when handling loose clothing such as skirts under large or challenging leg movements, LBS-based methods inevitably lead to severe structural distortions, unrealistic stretching, or cloth-body penetrations. In contrast, by decoupling the garment layer and integration with a physical simulator, our method \textcolor{black}{produces visually plausible cloth dynamics} effects and maintains structural integrity even under complex novel poses, as demonstrated in Figure~\ref{fig:lhm_comparison} and our supplementary video.

\subsection{Generated Data Visualization}
\begin{figure}[htbp]
  \centering
  \includegraphics[scale=0.35]{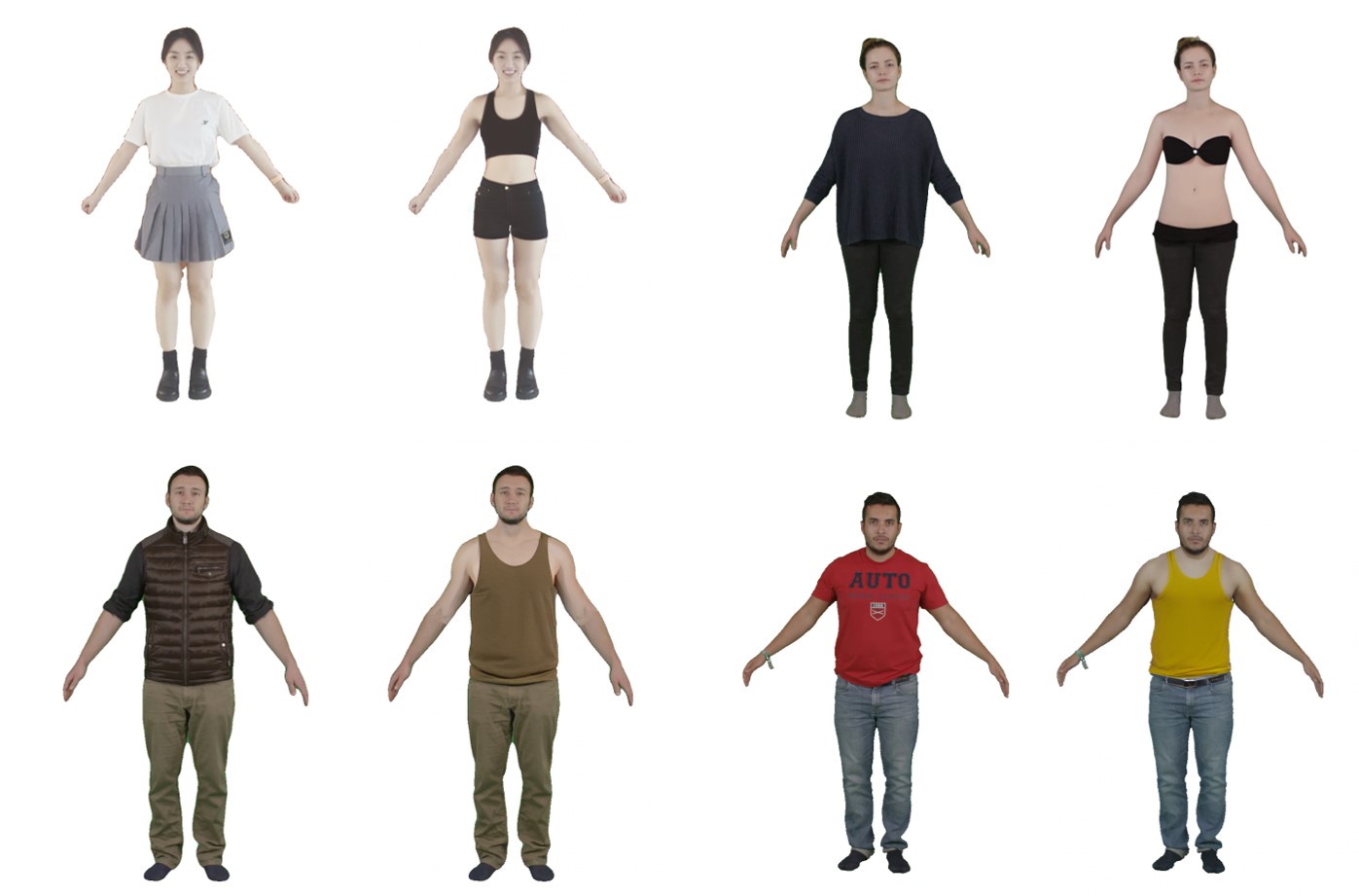}
  \begin{picture}(0,0)
    \put(-220,-8){Input}
    \put(-190,-8){Underlying Body}
    \put(-100,-8){Input}
    \put(-68,-8){Underlying Body}
  \end{picture}
  \vspace{3mm}
  \caption{Visualization of body generated data.
    }
  \label{body_gen}
\end{figure}
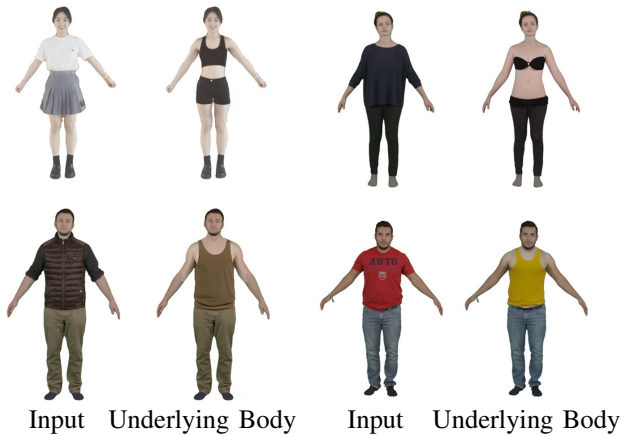

\begin{figure}[htbp]
  \centering
  \includegraphics[scale=0.25]{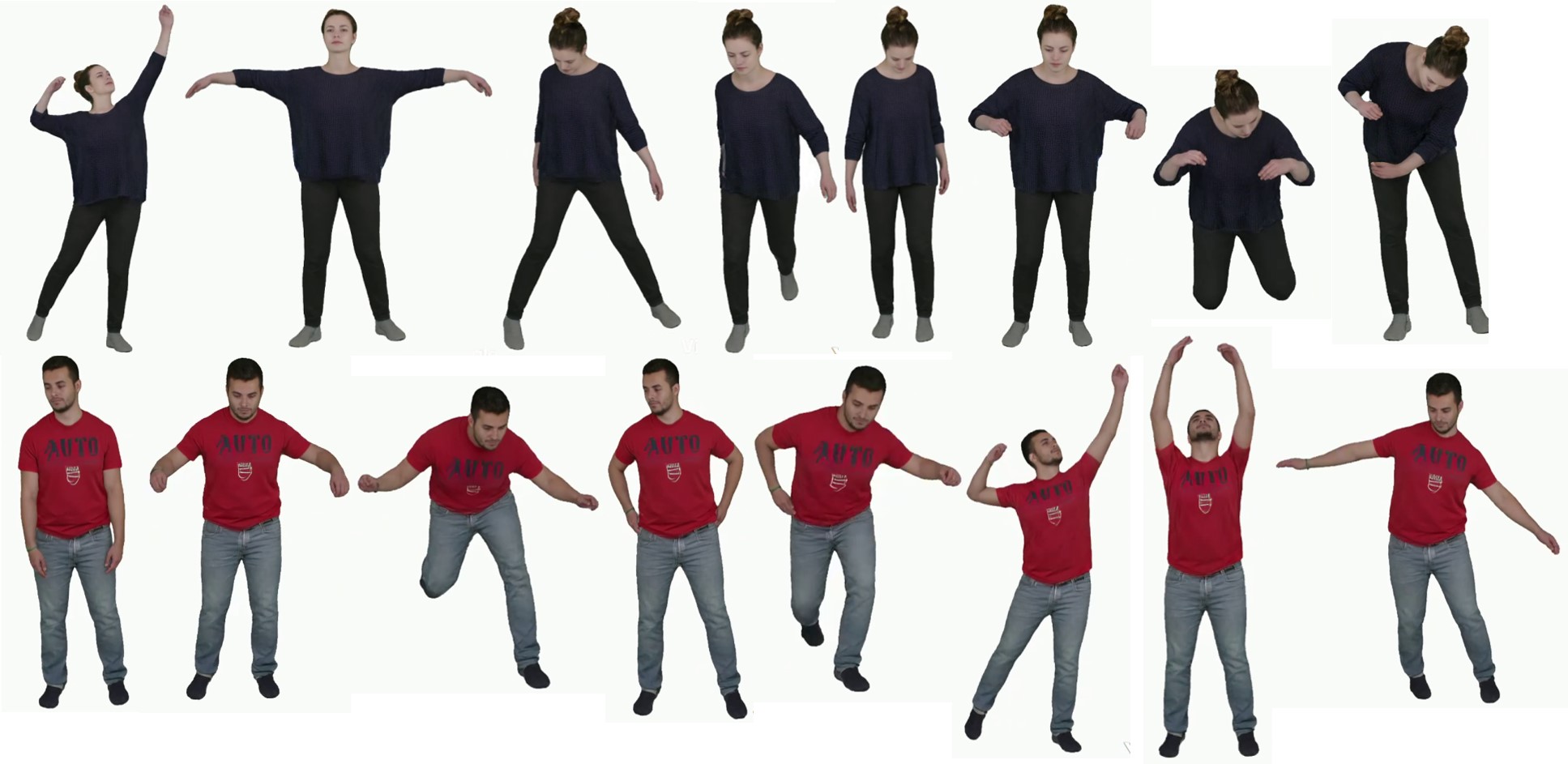}
  \begin{picture}(0,0)
    % \put(-300,-8){Input}
    % \put(-240,-8){Underlying Body}
    % \put(-130,-8){Input}
    % \put(-68,-8){Underlying Body}
  \end{picture}
  \caption{Visualization of key pose generated data.
    }
  \label{cloth_gen}
\end{figure}

% \textbf{Body Generated Data Visualization.}
In Figure~\ref{body_gen}, we present the results of generating images of underlying bodies from the raw input.
In Figure~\ref{cloth_gen}, we present the results of generating images of key pose.
The generated data for both body and clothing are reasonable and can compensate for missing information in monocular input.

\begin{figure}[htbp]
  \centering
  \includegraphics[scale=0.19]{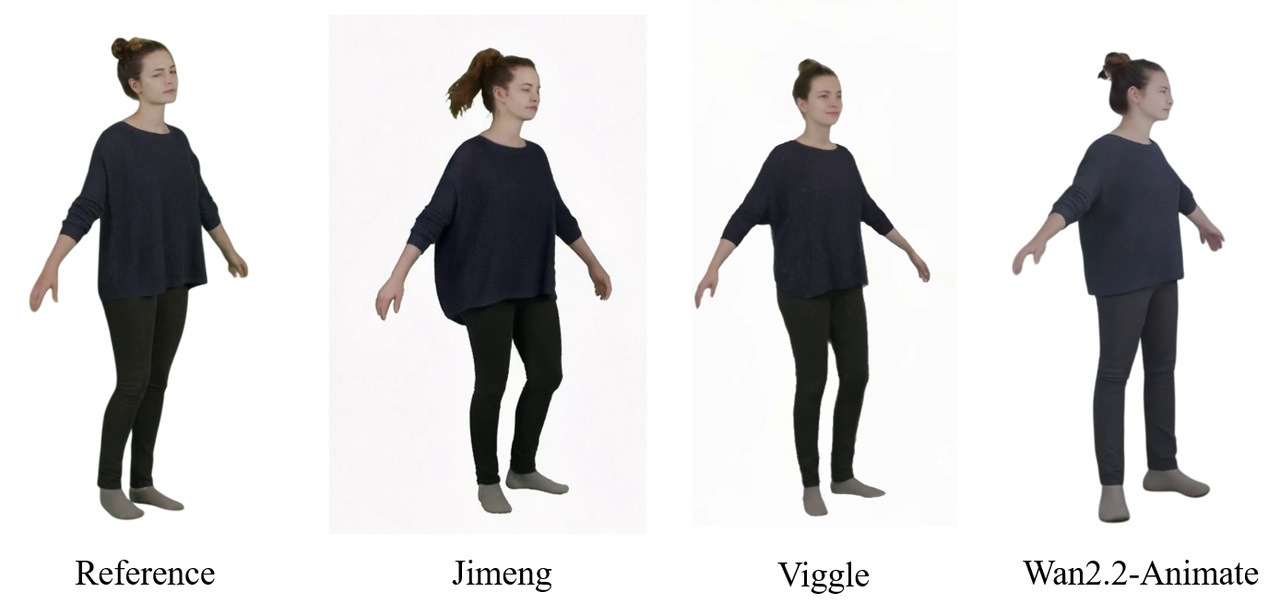}
  % \begin{picture}(0,0)
  %   \put(-230,-8){Input}
  %   \put(-190,-8){Underlying Body}
  %   \put(-100,-8){Input}
  %   \put(-68,-8){Underlying Body}
  % \end{picture}
  % \vspace{3mm}
  \caption{Visualization comparison of SOTA video generation models.
    }
  \label{gen_video_compare}
\end{figure}

In Figure~\ref{gen_video_compare}, we compared the commercial model Jimeng~\cite{jimeng2026} and the open-source models Wan2.2-Animate~\cite{cheng2025wananimateunifiedcharacteranimation} and Viggle~\cite{viggle2025}. Despite the rapid development of video generation models in recent years, they still suffer from temporal and spatial inconsistency issues, which also persist in human animation models. Jimeng fails to maintain consistency in hair details, Wan2.2-Animate exhibits overall color drift on clothing, while Viggle slightly alters the tightness and looseness of garments. In fact, it is difficult to determine which generation model is optimal, and selecting a specific generation model is not the core contribution of our work. The relatively robust Viggle was chosen for the experiments in this work. Our method can be directly migrated to any superior generation model once it becomes available.

\subsection{User Study on Novel Poses}
We selected 7 unseen sequences from AMASS~\cite{AMASS:ICCV:2019}, generated novel pose renderings using iHuman~\cite{pramishp2024iHuman} and our method, and conducted a user study.
% Users were asked to rate our method and iHuman on image clarity and dynamic clothing correctness for novel pose animation. Of the 50 valid responses, 97.1\% preferred our method in clarity, and 97.7\% rated it higher in dynamic clothing.
Users answered two questions: (1) Which video has higher clarity? (2) Which video demonstrates better accuracy in dynamic clothing? \textcolor{black}{The results are shown in} Table~\ref{tab_user_study_results}:
\begin{table}[htbp]
\centering  
\caption{User study results on visual clarity and dynamic clothing accuracy. A total of 49 valid questionnaires were collected.}
\vspace{-3mm}
\begin{tabular}{ccc}
\hline
                  & Visual Clarity & Dynamic Accuracy \\
 \hline                
iHuman            & 11             & 7               \\
\hline
\textbf{Ours}     & \textbf{332}   & \textbf{334}    \\
\hline
\end{tabular}
\label{tab_user_study_results}
\end{table}

A total of 49 valid questionnaires were collected.
\textcolor{black}{Each participant evaluated seven sequence pairs, yielding 343 valid judgments for visual clarity and 341 valid judgments for dynamic clothing accuracy. Our method was preferred in 332 of 343 judgments (96.79\%) for visual clarity and in 334 of 341 judgments (97.95\%) for dynamic clothing accuracy. These descriptive preference rates show that participants generally preferred our results over those of iHuman on the evaluated novel-pose sequences.}

\subsection{Ablation Study}
\begin{figure}[ht]
  \centering
  \includegraphics[width=\linewidth]{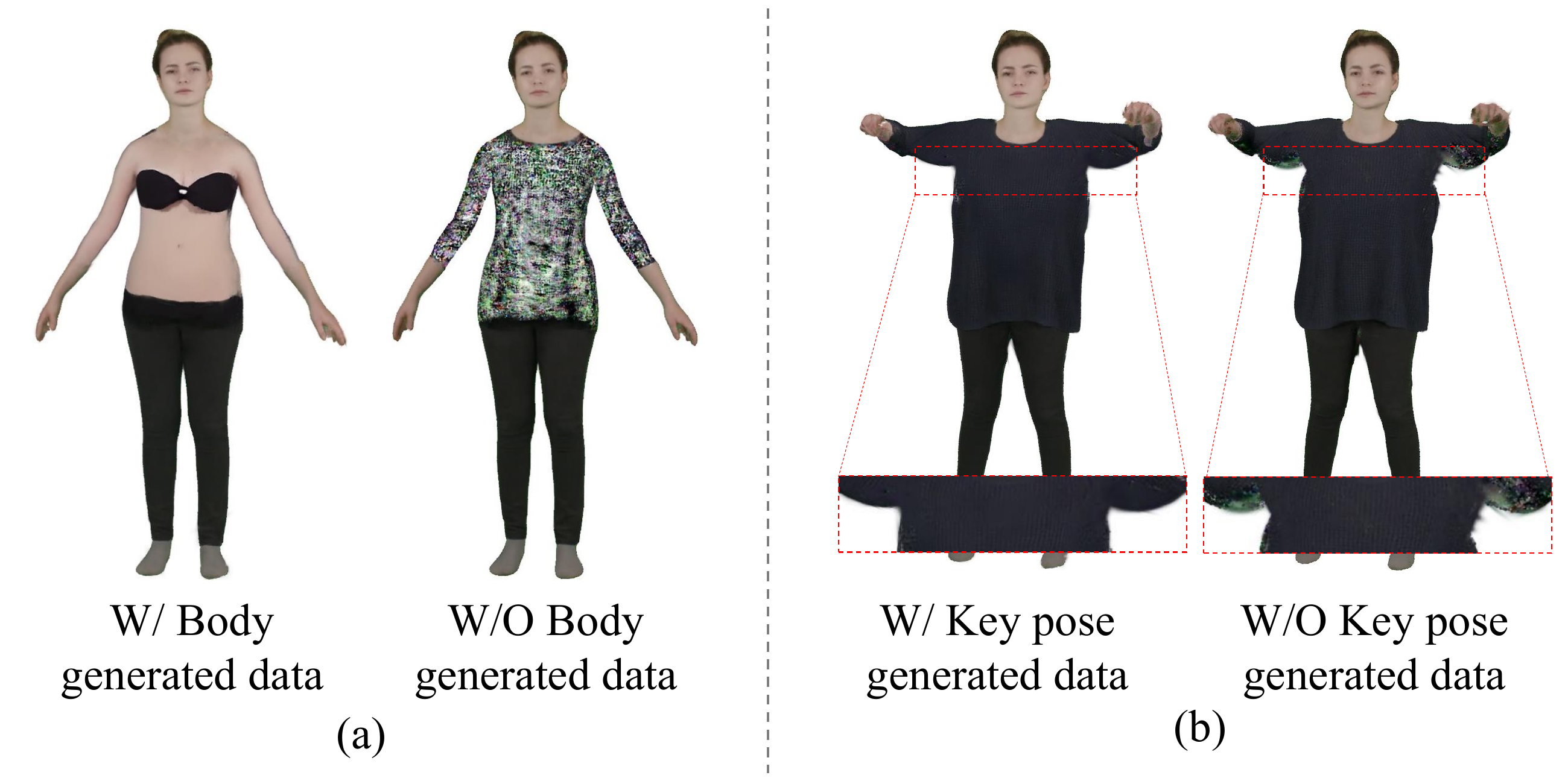}
  % \begin{picture}(0,0)
  %   \put(-175,-8){(a)}
  %   \put(-60,-8){(b)}
  % \end{picture}
  \vspace{-8mm}
  \caption{\textcolor{black}{Qualitative ablation of the two components of generated data.}(a) Ablation study on \textcolor{black}{removing underlying-body generated data.}  
  (b) Ablation study on \textcolor{black}{removing key-pose generated clothing data.}
    }
  \label{ablation_data_fig}
\end{figure}

\textbf{Body Data Generation.}
We attempted to remove the body-generated data and render the body independently to evaluate its effect.
In Figure~\ref{ablation_data_fig} (a), we compare the results with and without using body generated data. 
Without body-generated data, artifacts are evident in the exposed regions. In contrast, incorporating body-generated data for supervision provides supplementary information for these unseen regions, effectively reducing artifacts and improving the overall reconstruction quality.

\textbf{Key Pose Data Generation.}
To evaluate the impact of key pose generated data, we conducted an experiment comparing results with and without this data, as shown in Figure~\ref{ablation_data_fig} (b). In the absence of key pose generated data, occluded regions in the input, such as the underarm area, result in noticeable artifacts when animating the reconstructed human avatar to new poses. Conversely, incorporating  key pose generated data effectively fills in the appearance details for these occluded regions, reducing artifacts and significantly improving the overall reconstruction quality. This demonstrates the importance of key pose generated data in enhancing the realism and animated results.

\begin{figure}[htbp]
  \centering
  \includegraphics[scale=0.27]{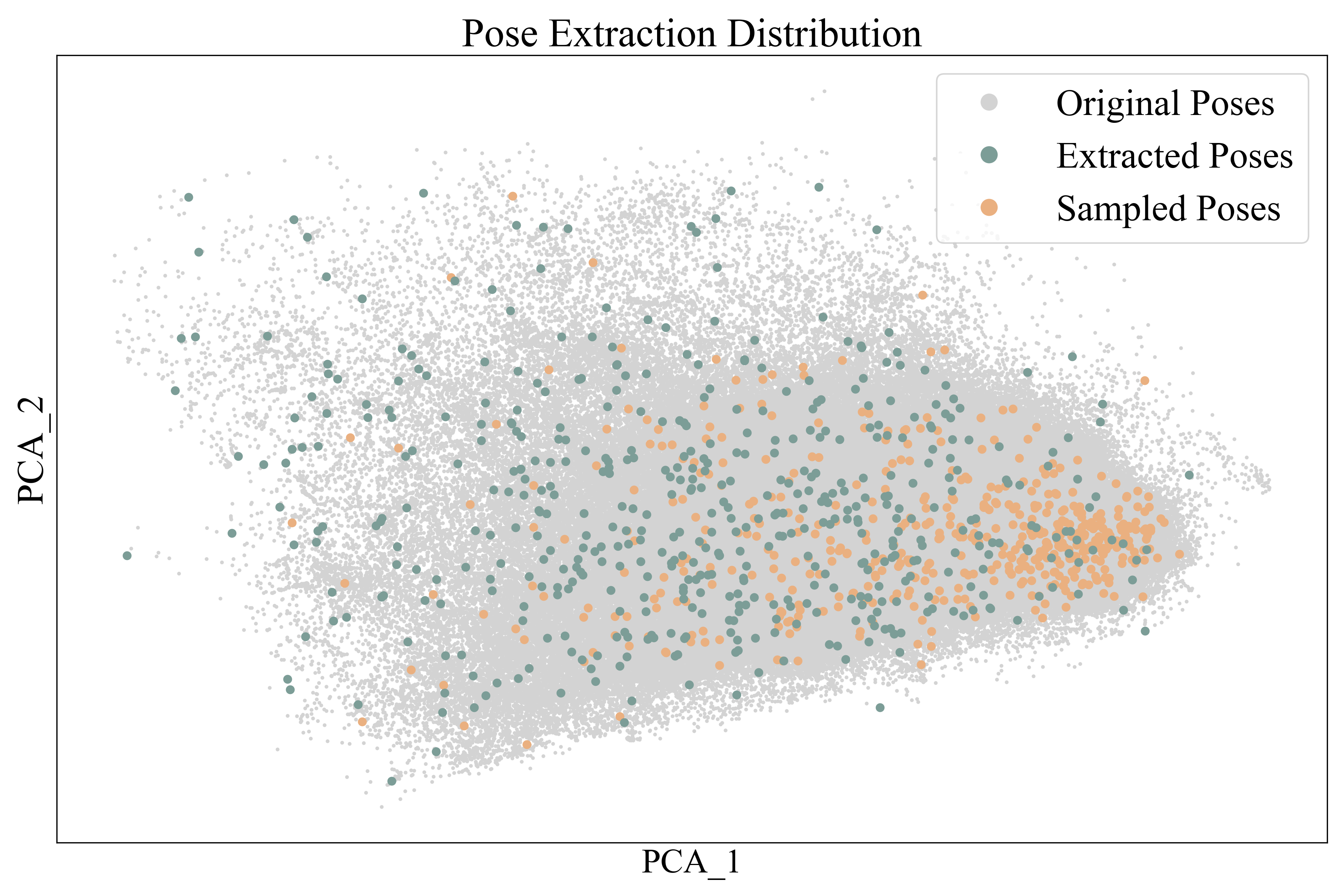}
  \begin{picture}(0,0)
    % \put(-175,-8){(a)}
    % \put(-60,-8){(b)}
  \end{picture}
  % \vspace{-5mm}
  \caption{Ablation Study on Keyframe Extraction Using MMR.
  ``Original Poses'' refers to the original large-scale pose data, ``Extracted Pose'' refers to the pose set we extracted, and ``Sampled Poses'' refers to the poses obtained using the uniform sampling strategy.
    }
  \label{mmr}
\end{figure}

\textbf{Keyframe Extraction Using MMR.}
We use the MMR algorithm to extract a key pose set by considering pose representativeness, relevance, and visibility. In this experiment, we compare our approach with uniform sampling strategy to demonstrate the effectiveness of the extracted key pose set, as shown in Figure ~\ref{mmr}.
For uniform sampling, extracting one frame at fixed intervals tends to concentrate the selected frames in regions where the original pose distribution is dense, which is not conducive to capturing more diverse information. 
In contrast, the MMR pose method produces a more evenly distributed pose set, facilitating the generation of data with richer and more comprehensive pose information.

 To evaluate the quality of the key pose generating data, we designed the following experiment: For each subject, we used the key pose data and only 10-frame GT data to train, and tested on the rest of the GT data. We generated 200-frame data by our pose MMR module, and 200, 500, 1000-frame by \textcolor{black}{uniform sampling} for comparison. We also trained without any generating data as an ablation on using of the key pose data. The quantitative results are in Table~\ref{mmr_random}.
 
 The quality of 200-frame data sampled by pose MMR is slightly lower than that of 1000-frame data sampled uniformly, indicating that our pose MMR strategy can make the generating data contain more information in fewer frames. And this key pose data can fill in some of the missing information in the GT data.
\begin{table}[htbp]
\centering 
\caption{Quantitative Comparison of 200 MMR samples, 200 \textcolor{black}{uniform} samples, 500 \textcolor{black}{uniform} samples, and 1000 \textcolor{black}{uniform} samples.}
\vspace{-3mm}
\begin{tabular}{cccc}
\hline
                                 & PSNR     & SSIM     & LPIPS    \\
 \hline               
200 MMR                          & 28.30 & 0.9722 & 0.0417 \\
\hline
200 \textcolor{black}{Uniform}                       & 25.33    & 0.9348   & 0.0714   \\ 
\hline
500 \textcolor{black}{Uniform}                       & 27.58    & 0.9699   & 0.0489   \\ 
\hline
1000 \textcolor{black}{Uniform}                      & 28.69    & 0.9737   & 0.0410  \\
\hline
\end{tabular}
\label{mmr_random}
\end{table}

\begin{figure}[ht]
  \centering
  \includegraphics[width=\linewidth]{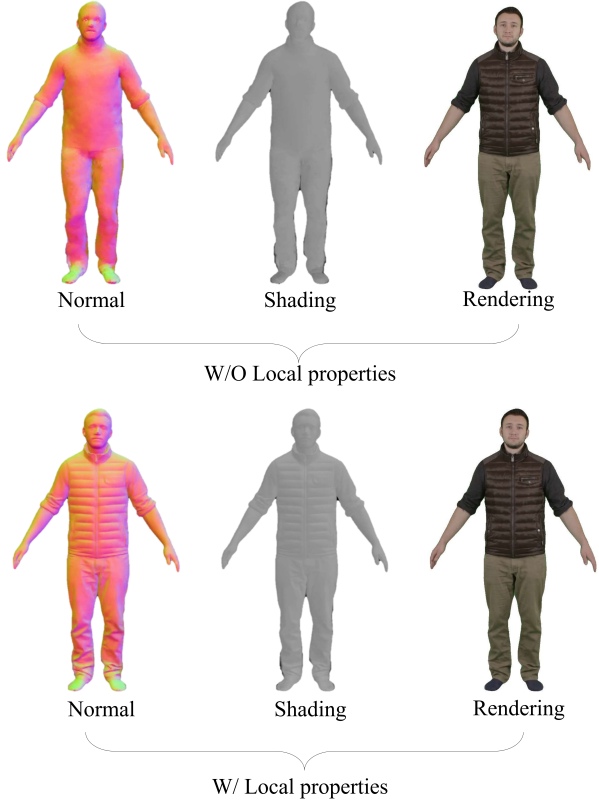}
  \begin{picture}(0,0)
    % \put(-175,-8){(a)}
    % \put(-60,-8){(b)}
  \end{picture}
  \vspace{-5mm}
  \caption{Ablation study on local properties.
    }
  \label{ablation_local}
\end{figure}

\textbf{Learnable Local Properties.}
For each 2DGS, we augmented learnable local geometric properties to enhance its expressiveness.
In this experiment, we compared rendering effects of our representations with and without learnable local properties.
As shown in the Figure~\ref{ablation_local}, results with learnable local properties exhibit richer geometric and shading details.

\textbf{Generated Data Alignment.}
We perform an ablation study on the proposed generation data alignment method, focusing on shape consistency, style consistency, pose alignment, and SH coefficients. Quantitative results show that removing any of these components leads to a decline in reconstruction quality, as shown in Table~\ref{generated_data_alignment}.

\begin{table}[hbp]
\centering 
\caption{Quantitative ablation study on generated data alignment strategy.}
\begin{tabular}{cccc}
\hline
Method                            & PSNR     & SSIM     & LPIPS    \\
 \hline               
w/o Shape Consistency & 31.16 & 0.9848 & 0.0242 \\
\hline
w/o Style Consistency & 31.85 & 0.9869 & 0.0209 \\
\hline
w/o Pose Alignment & 31.38 & 0.9854 & 0.0227 \\
\hline
w/o SH Coefficients & 31.93 & 0.9871 & 0.0206 \\
\hline
Full & \textbf{32.49} & \textbf{0.9876} & \textbf{0.0198} \\
\hline
\end{tabular}
\label{generated_data_alignment}
\end{table}

\begin{table}[htbp]
\centering 
\caption{Quantitative ablation on core components and comparison with PICA variants on the f3c subject. We evaluate the impact of integrating our proposed architecture (SH \& Local properties) and \textcolor{black}{Generated} Data into both the baseline PICA and our framework.}
\vspace{-3mm}
\begin{tabular}{lccc}
\hline
Method                             & PSNR$\uparrow$ & SSIM$\uparrow$ & LPIPS$\downarrow$ \\
\hline                
PICA (Baseline)                    & 30.92          & 0.9795         & 0.0265            \\
PICA + \textcolor{black}{Generated} Data                   & 29.05          & 0.9732         & 0.0286            \\
PICA + SH \& Local                 & 28.67          & 0.9757         & 0.0196            \\
PICA + SH \& Local + \textcolor{black}{Generated} Data     & 27.80          & 0.9730         & 0.0240            \\
\hline
Ours w/o Local Properties          & 31.89          & 0.9859         & 0.0216            \\
Ours w/o SH                        & 33.49          & 0.9874         & 0.0245            \\
Ours w/o \textcolor{black}{Generated} Data                 & \textbf{33.67} & \textbf{0.9903} & \textbf{0.0164}   \\
Ours (Full Pipeline)               & 32.49          & 0.9876         & 0.0198            \\
\hline
\end{tabular}
\label{tab_rep_vs_data}
\end{table}

\textbf{Decoupling Representation and Generated Data.}
To address potential concerns regarding the reliance on generated data and \textcolor{black}{better understand the respective contributions of our reconstruction framework and Generated Data}, we conduct a comprehensive decoupling experiment on the f3c subject. As shown in Table~\ref{tab_rep_vs_data}, \textcolor{black}{ours w/o Generated Data achieves a PSNR of 33.67, compared with 30.92 for the baseline PICA, while neither setting uses Generated Data. These results indicate that the performance gain cannot be attributed solely to generated-data augmentation.}
% This confirms that the fundamental performance improvement stems from our novel 3D representation rather than data augmentation.
In fact, directly introducing Generated Data into PICA degrades reconstruction metrics, suggesting that generated observations may act as noisy supervision when their inconsistencies are not properly handled. The Ours w/o Generated Data variant achieves better reconstruction metrics on held-out frames from the captured sequence, whereas the role of Generated Data is mainly to provide complementary supervision for insufficiently observed regions, which becomes important under novel poses, as qualitatively demonstrated in Fig~\ref{ablation_data_fig}.

More importantly, we integrate our generated data \textcolor{black}{(Generated Data)} and architectural components into the PICA baseline. Surprisingly, adding \textcolor{black}{generated data} to PICA degrades its PSNR from 30.92 to 29.05. This occurs because generated data inherently contains geometric and appearance inconsistencies across frames. Without our proposed alignment strategies and dual-layer formulation, these inconsistencies act as severe noise, corrupting the optimization of baseline methods. Our framework, in contrast, successfully mitigates these artifacts, making the generative priors usable.

Finally, comparing our full pipeline with our \textcolor{black}{\textit{w/o Generated Data}} variant, we observe a deliberate and expected trade-off: incorporating \textcolor{black}{generated data} slightly lowers the strictly overfitted reconstruction metrics on \textcolor{black}{held-out frames}, but \textcolor{black}{it provides important complementary supervision for synthesizing plausible appearances in previously unseen regions} during novel pose animation, as demonstrated in our qualitative evaluations.
\textcolor{black}{This trade-off is also qualitatively illustrated in Fig.~\ref{ablation_data_fig}, where removing the generated data leads to more visible artifacts in previously unseen or occluded regions under novel poses.}

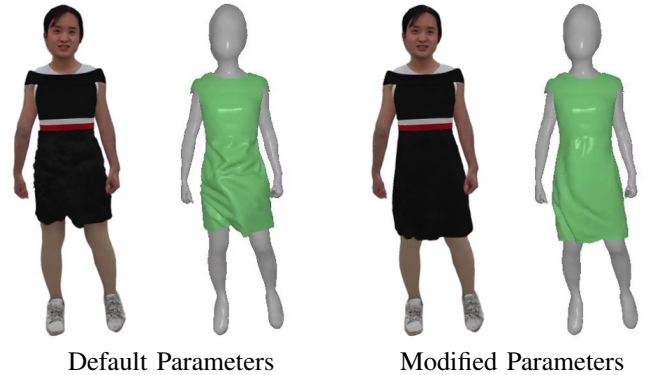
\begin{figure}[htbp]
  \centering
  \includegraphics[width=\linewidth]{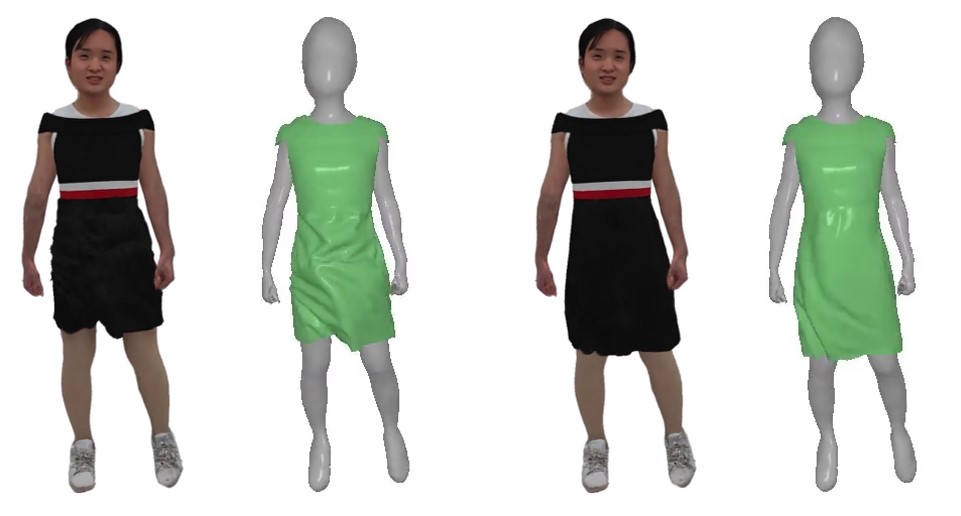}
  \vspace{-5mm}
    \begin{picture}(0,0)
    \put(-100,5){Default Parameters}
    \put(25,5){Modified Parameters}
    % \put(-130,-8){Input}
    % \put(-68,-8){Underlying Body}
  \end{picture}
  \caption{Comparison of garment dynamics driven by optimized parameters versus default simulator parameters.}
  \label{fig:default_physics}
\end{figure}

\textbf{Effectiveness of Physical Parameter Optimization.}
Our framework utilizes a pre-trained GNN-based simulator~\cite{grigorev2024contourcraft}, which inherently possesses strong physical priors and can generate reasonable cloth dynamics even with default parameters. Therefore, the goal of our physics fitting stage is not to learn properties from scratch, but rather to fine-tune these default parameters to better approximate the specific fabric characteristics observed in the monocular video. 
As shown in Figure~\ref{fig:default_physics}, compared to the generic deformations produced by the default parameters, our refined parameters yield dynamic behaviors that more closely align with the specific garment. Note that perfectly recovering absolute physical properties from highly under-constrained monocular video is intractable. Our optimization serves as an effective refinement over the strong GNN prior.

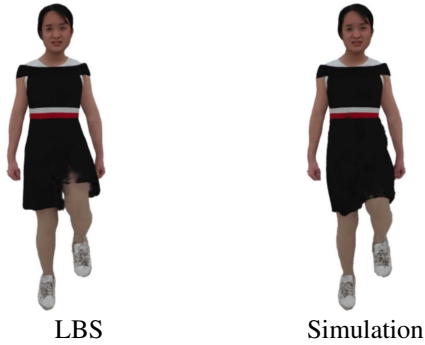
\begin{figure}[htbp]
  \centering
  \includegraphics[scale=0.33]{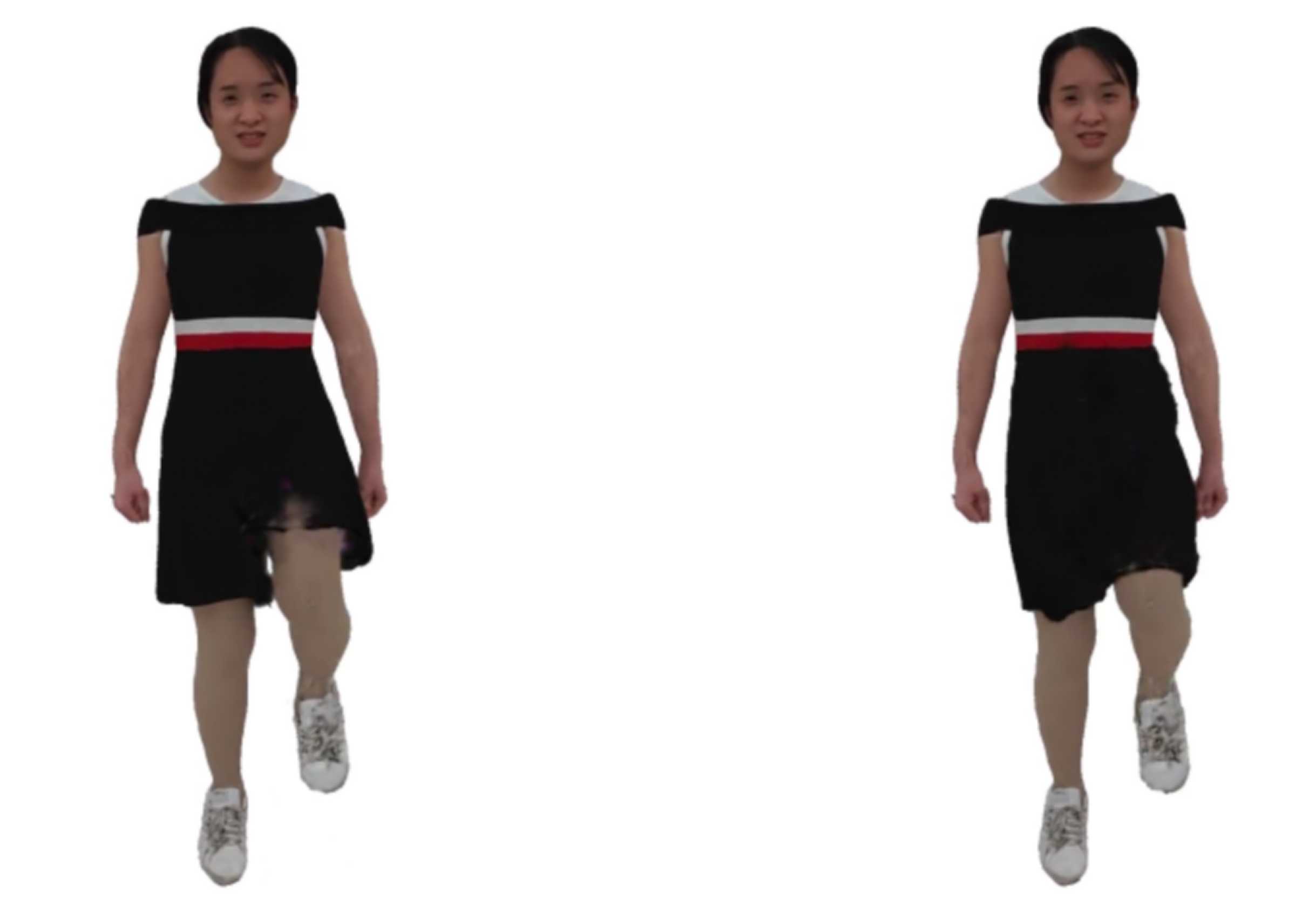}
  \vspace{-5mm}
  \begin{picture}(0,0)
    \put(-147,-8){LBS}
    \put(-52,-8){Simulation}
    % \put(40,5){LBS}
    % \put(80,5){Simulation}   
    % \put(-130,130){(a)}
    % \put(20,130){(b)}  
    % \put(-130,-8){Input}
    % \put(-68,-8){Underlying Body}
  \end{picture}
  \vspace{5mm}
  \caption{Ablation on animation strategy. Comparison between conventional LBS and physics-integrated simulation. }
  \label{fig:ablation_physics_geo}
\end{figure}

\textbf{Physics-based Animation and Geometric Validation.}
To demonstrate the necessity of our physics-integrated animation, we compare our full pipeline against an internal baseline that replaces the physics simulation with traditional Linear Blend Skinning (LBS) for driving the reconstructed mesh. As shown in Figure~\ref{fig:ablation_physics_geo}, when applying LBS to loose clothing such as a skirt, the garment mesh strictly follows the leg bones, leading to severe unnatural stretching during wide stances or novel motions. Our physics simulation effectively resolves this by modeling collisions and non-rigid dynamics. 
% Furthermore, to explicitly verify that these realistic dynamic effects stem from actual 3D geometric deformations rather than appearance-level texture illusions, we visualize the skirt rendering without texture (plain shading) in Figure~\ref{fig:ablation_physics_geo} (b). The texture-less rendering clearly confirms that the dynamic wrinkles and cloth swings are intrinsically captured by our underlying mesh geometry.

\subsection{Applications}
\label{apply}

\subsubsection{Virtual Try-On}
We demonstrate the performance of our reconstructed human avatar in virtual try-on in Figure~\ref{tryon}, along with a comparison to the 2D virtual try-on method IDM-VTON~\cite{choi2024idmvton}.
Since IDM-VTON requires specifying the body for each frame, we demonstrate the try-on results only based on real captured data.
For clothing, we use front-facing images as input.
IDM-VTON can preserve texture features during try-on, but it sometimes alters the style of the clothing and fails to ensure frame-to-frame consistency. Additionally, it cannot perform try-on in novel poses.
Our method achieves virtual try-on through clothing transfer and the integration of physical simulation, 
and preserves the texture and style of the clothing during try-on, \textcolor{black}{and produces visually plausible cloth dynamics}.
Our method can perform try-on with clothing reconstructed from monocular input.
Please see our video for more try-on results.

\subsubsection{Relighting}
\begin{figure}[htbp]
  \centering
  \includegraphics[width=\linewidth]{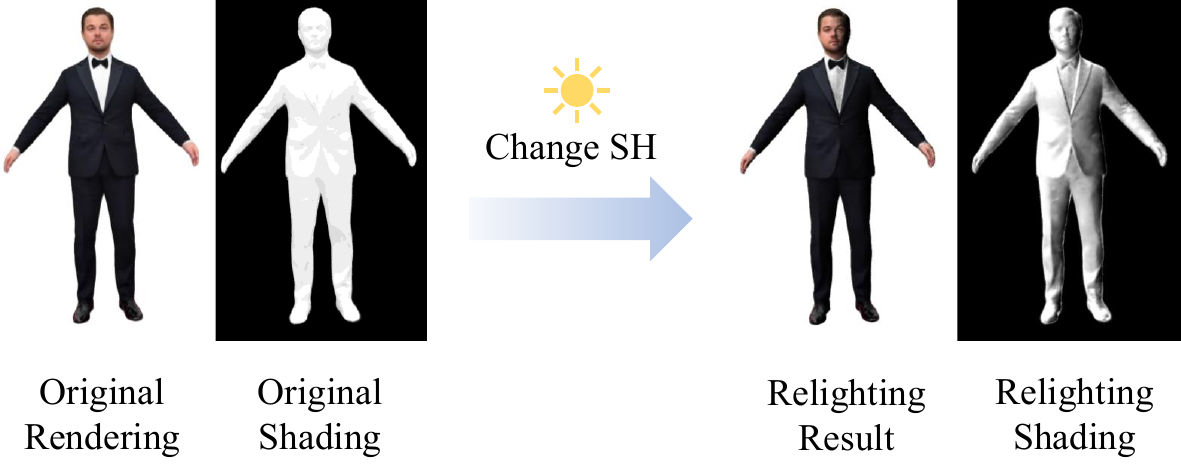}
  \begin{picture}(0,0)
    % \put(-175,-8){(a)}
    % \put(-60,-8){(b)}
  \end{picture}
  \vspace{-7mm}
  \caption{Relighting via SH Coefficient manipulation.}
  \label{relighting}
\end{figure}

Our reconstructed human avatars can achieve relighting effects during rendering by modifying SH coefficients. As shown in Figure~\ref{relighting}, the Relighting Shading changes with adjustments to SH coefficients, which consequently produces relighting effects in the rendered color images.
% \begin{figure}[ht]
%   \centering
%   \includegraphics[scale=0.4]{Figs/Figs_asia/relighting.pdf}
%   \begin{picture}(0,0)
%     % \put(-175,-8){(a)}
%     % \put(-60,-8){(b)}
%   \end{picture}
%   \vspace{-8mm}
%   \caption{Relighting via SH Coefficient manipulation.
%     }
%   \label{relighting}
% \end{figure}

\vspace{-7mm}
\begin{figure}[htbp]
  \centering
  \includegraphics[width=\linewidth]{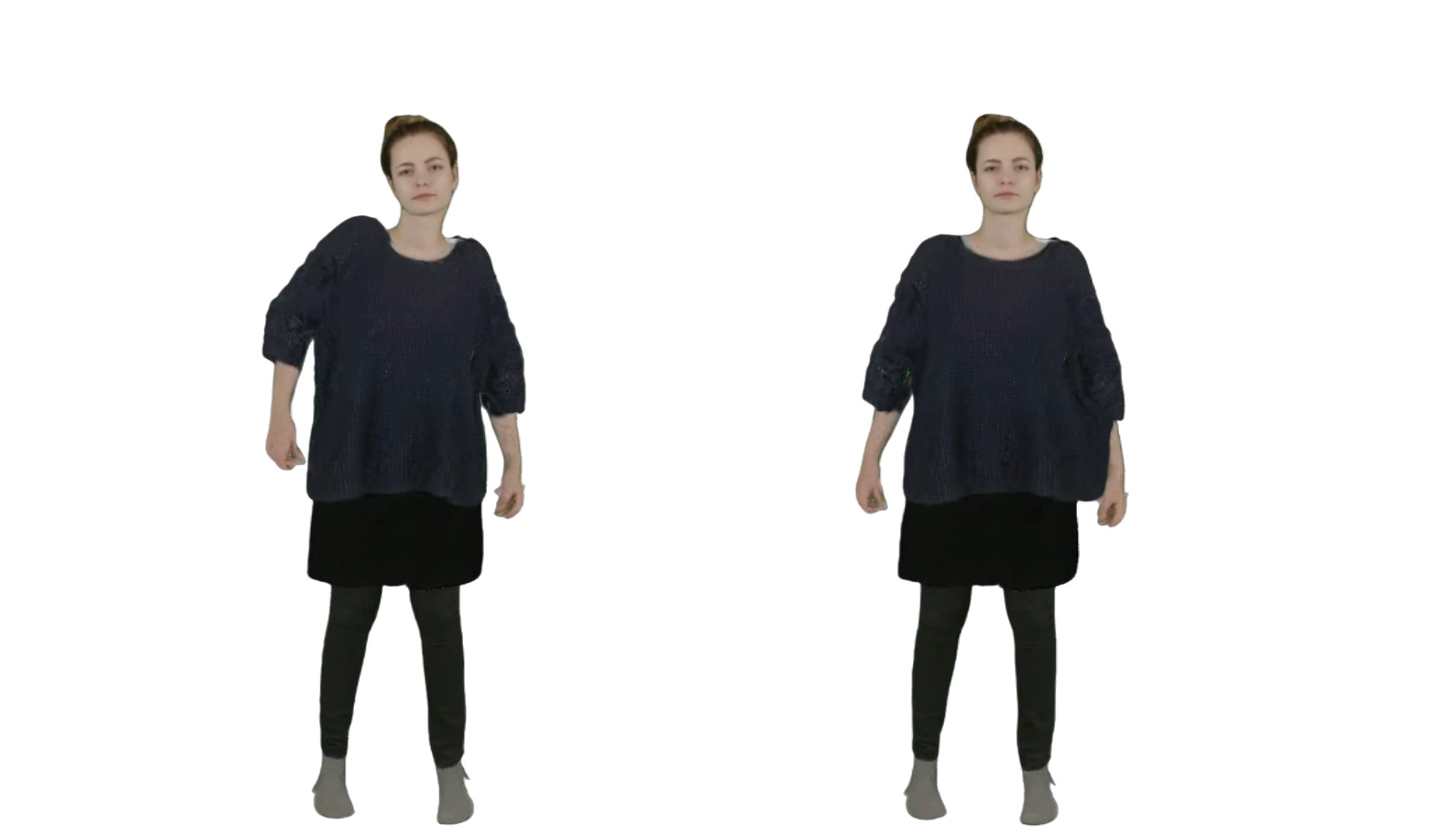}
  \begin{picture}(0,0)
    % \put(-175,-8){(a)}
    % \put(-60,-8){(b)}
  \end{picture}
  \vspace{-7mm}
  \caption{Multi-layer clothing driving demonstration. Garment transfer from chh to f3c.}
  \label{fail_case}
\end{figure}

\vspace{2mm}

% \subsection{Failure Cases.}

\subsection{Limitations and Future Work.}
\textcolor{black}{Our method relies on several off-the-shelf priors for initialization. Severe errors in body/garment geometry, pose estimation, segmentation, or generated unseen regions may propagate to the final reconstruction. In particular, our alignment strategy can alleviate inconsistencies across generated observations, but cannot fully correct semantically incorrect hallucinations in completely unseen regions. Such errors may further affect garment-body collision and physical simulation.}

Although our method \textcolor{black}{optimizes} the physical parameters from monocular videos, its fidelity cannot always be guaranteed due to the following reasons:
(1) Most self-supervised neural cloth simulation methods including ContourCraft~\cite{grigorev2024contourcraft} focus on reasonable forward inference rather than estimating inverse parameters based on real-world data.
(2) There is a gap between the template geometry reconstructed from monocular input and the ground truth geometry, which may result in errors during collision calculations.
(3) The wrinkling of clothing is not only determined by the body moving, inherently exhibit uncertainty.
% High-fidelity animation rendering depends not only on physical parameters but also on the interplay of geometry, appearance, animation, and environment.
% The integrated physical simulation enables high-fidelity animation effects for the geometry.
For appearance, although we use generated data to enhance its generalization, generative models may not always produce high-quality results for all clothed human inputs.
% , particularly for challenging poses, uncommon clothing, and maintaining identity consistency throughout sequences.
% Additionally, our method does not account for environmental lighting, which can lead to inconsistencies during virtual try-on if the lighting conditions for the clothing and body are inconsistent.
% This might be solved by introducing the relighting module of Animatable Gaussians~\cite{li2024animatablegaussians}, which we leave as a future work.
(4) While our model demonstrates remarkable efficacy in both try-on and novel pose synthesis tasks, its extension to multi-layered garment animation introduces extra complexities. As illustrated in Figure~\ref{fail_case}, evaluating the framework on sophisticated multi-layered apparel can occasionally introduce unresolved cross-layer interactions during dynamic driving, which may subsequently yield subtle geometric inter-penetration and rendering artifacts. Addressing these complex layered configurations without relying on explicit physics engines remains an open challenge that warrants further investigation.

\section{Conclusion}

We proposed a novel method capable of reconstructing decoupled clothed human avatars from monocular inputs and \textcolor{black}{enabling virtual try-on through physical simulation with visually plausible cloth dynamics}. 
By combining input real video data with generated data, our method can reconstruct a double-layer Mesh-GS representation of the human body with strong generalization capabilities for novel poses and clothing, integrating physical simulation. 
For the inter-frame inconsistency in generated data, we decompose it into geometric inconsistency and appearance inconsistency and address them separately.

Thanks to this innovative approach, we can achieve human avatar reconstruction for virtual try-on with only monocular input, \textcolor{black}{delivering high-fidelity results with visually plausible cloth dynamics for novel garments and poses}. 
Our method requires only a single camera to achieve 3D physics-integrated reconstruction, providing a technical foundation for animation and virtual try-on applications.

\IEEEpeerreviewmaketitle

\ifCLASSOPTIONcompsoc
  \section*{Acknowledgments}
\else
  \section*{Acknowledgment}
\fi

This research was supported by the National Natural Science Foundation of China (No.62272433), Anhui Provincial Natural Science Foundation (No.2508085ZD011), Key Science \& Technology Project of Anhui Province (No. 202523o09050004) and the Fundamental Research Funds for the Central Universities.

\ifCLASSOPTIONcaptionsoff
  \newpage
\fi

\bibliographystyle{IEEEtran}
\bibliography{main}

\end{document}